Conflict in Africa during COVID-19: social distancing, food vulnerability and welfare response

Roxana Gutiérrez-Romero♣

28 MAY 2020

ABSTRACT

We study the effect of social distancing, food vulnerability, welfare and labour COVID-19 policy responses on riots, violence against civilians and food-related conflicts. Our analysis uses georeferenced data for 24 African countries with monthly local prices and real-time conflict data reported in the Armed Conflict Location and Event Data Project (ACLED) from January 2015 until early May 2020. Lockdowns and recent welfare policies have been implemented in light of COVID-19, but in some contexts also likely in response to ongoing conflicts. To mitigate the potential risk of endogeneity, we use instrumental variables. We exploit the exogeneity of global commodity prices, and three variables that increase the risk of COVID-19 and efficiency in response such as countries colonial heritage, male mortality rate attributed to air pollution and prevalence of diabetes in adults. We find that the probability of experiencing riots, violence against civilians, food-related conflicts and food looting has increased since lockdowns. Food vulnerability has been a contributing factor. A 10% increase in the local price index is associated with an increase of 0.7 percentage points in violence against civilians. Nonetheless, for every additional anti-poverty measure implemented in response to COVID-19 the probability of experiencing violence against civilians, riots and food-related conflicts declines by approximately 0.2 percentage points. These anti-poverty measures also reduce the number of fatalities associated with these conflicts. Overall, our findings reveal that food vulnerability has increased conflict risks, but also offer an optimistic view of the importance of the state in providing an extensive welfare safety net.

♣ Centre for Globalisation Research (GCR) working paper series, School of Business and Management, Queen Mary University of London, London, UK. r.gutierrez@qmul.ac.uk.

1. INTRODUCTION

In response to the COVID-19 pandemic, several governments have implemented social distancing measures. Although these measures have shown to be effective in curbing the spread of the novel-coronavirus, they have also shown to cause significant economic, social and political disruption (Barrett, 2020; Senghore, Savi, Gnangnon, Hanage, & Okeke, 2020). This is particularly the case for the developing world, which relies largely on the informal economy, still has high levels of poverty, with weak health and welfare systems, and where the majority of the population simply do not have the luxury to work remotely. For instance, in Africa, right before the pandemic outbreak, one in every five people was suffering from severe food insecurity, affecting nearly 277 million people. These vulnerable people had run out of food, most likely experienced hunger, even gone for days without eating, putting their well-being at a great danger (FAO, IFAD, UNICEF, WFP, & WHO, 2019). As a result of the pandemic, several forecasts predict that between 60-240 million people worldwide could be pushed into poverty, depending on the efficiency in providing urgent and adequate relief to vulnerable citizens and struggling businesses (Ahamed & Gutiérrez-Romero, 2020; Sumner, Hoy, & Ortiz-Juarez, 2020). The sudden loss of jobs and livelihoods for millions of people have caused food shortages and inflation, an explosive combination for uprisings.

This paper analyses two key questions. First, to what extent are social distancing measures, lockdowns and food vulnerability fuelling conflicts. Second, whether recently implemented COVID-19 anti-poverty programmes could curb such conflicts. We focus on the 24 African countries for which we have monthly data on local prices and real-time conflict data reported in the Armed Conflict Location and Event Data Project (ACLED).[1] We analyse four types of conflicts likely to arise as a result of COVID-19: riots, violence against civilians, food-related conflicts and food looting. We focus on these events from 1 January 2015 until 2 May 2020. We combine ACLED with the exact dates of early social distancing and lockdowns. To assess the role of food vulnerability and conflict, we construct a monthly index of local prices based on data from the Global Food Prices Database (WFP) and the USAID FEWS-NET. These datasets provide monthly commodity prices at a sub-country level, across 990 local markets, in 24 African countries for the entire period analysed here. We also construct an index of welfare and labour COVID-19 policy response based on the 12 types of interventions (such as cash-based transfers, utility support and wage subsidy), gathered by (Gentilini, Almenfi, Dale, Demarco, & Santos, 2020). We take the date of implementation of these measures from Hale et al. (2020). We add a wide-range of georeferenced controls at the sub-country level for areas of approximately 55x55 km including nightlight, mobile phone coverage of 2G-3G, percentage of mountains, the existence of petroleum fields, mines, diamond mines, size of the area, electricity coverage, primary roads coverage, population, infant mortality rate and cultivated land.

---

[1] ACLED provides real-time georeferenced data (with latitude and longitude coordinates) on the number of conflicts, associated fatalities, actors involved and exact date, including day and month of conflict (C. Raleigh & Dowd, 2016).

The man-imposed mobility restrictions to curb the COVID-19 pandemic add an extra layer of complexity to ongoing conflicts and food vulnerability. COVID-19 interventions (social distancing, welfare and labour policies) have been strongly dependent on political, economic and social contexts, thus, they are unlikely to be exogenous to existing conflicts. To mitigate potential endogeneity concerns, we use instrumental variables. As instruments we use the male mortality rate attributed to household and ambient air pollution for the year 2016 and the percentage of diabetes prevalence among the adult population (aged 20 -79) over the years 2010-2019. Both instruments are known risk factors to COVID-19 mortality (Fattorini & Regoli, 2020; Hussain, Bhowmik, & do Vale Moreira, 2020), and are thus likely to influence the decision of the state as when to impose social distancing measures. We also use the IMF global commodity monthly price index as a proxy for exogenous economic shocks. We also consider as instrument the colonial heritage of the analysed countries, as colonial history is known to affect the quality of existing institutions (Nash & Patel, 2019).

The paper offers four key findings. First, there is no evidence that early social distancing measures, such as banning some international flights, fuelled conflicts. However, despite the global call for ceasefire during the pandemic, local lockdowns have increased the probability of countries experiencing riots, violence against civilians and food-related conflicts. Second, we find that a 10% increase in the local price index is associated with a 0.7 percentage point increase in violence against civilians. This violence is more likely to occur in areas with cultivated land, in agreement with the theoretical literature that suggests that when food supply declines these areas are more vulnerable to rebel groups seeking resource appropriation, such as food (Rezaeedaryakenari, Landis, & Thies, 2020). Third, we find that the urgent welfare and labour anti-poverty initiatives implemented in light of COVID-19 have contributed to reducing the conflicts analysed. For instance, for every additional anti-poverty measure (nearly a 0.1 increase in the welfare/labour COVID-19 index), the probability of experiencing violence against civilians, riots and food-related conflicts declines by approximately 0.2 percentage points. These anti-poverty measures also reduce the number of fatalities associated with these conflicts. Fourth, we also analysed the number of conflicts in which the state was directly involved as an actor (either instigating or responding to contain violence) and distinguish between food related and violence against civilians. We find that in countries that have provided a higher number of welfare and labour anti-poverty policies, the state is less likely to be involved as an actor in food-related conflicts. Paradoxically, in these countries the state is more likely to be involved as an actor in violence against civilians, but the evidence suggests this is an attempt to strictly enforce local lockdowns.

There is scant but growing literature on the relationship between aid, anti-poverty projects and conflict (E. Berman, Shapiro, & Felter, 2011; Crost, Felter, & Johnston, 2014; Nunn & Qian, 2014). The literature has offered quite mixed findings and is far from reaching a consensus. Nonetheless, there is more promising evidence that cash transfers are successful in reducing food vulnerability and poverty in the Africa context (Chakrabarti, Handa, Natali, Seidenfeld, & Tembo, 2020). There is also evidence that (conditional) cash transfers can reduce the incidence of violent conflicts if adequately tailored to local contexts (Crost, Felter, & Johnston, 2016; Pena, Urrego, & Villa, 2017). Our results resonate with these encouraging findings on conflict reduction.

A wide range of anti-poverty policies has been implemented (with at least five simultaneous and ongoing anti-poverty COVID-19 initiatives in the most active countries analysed here). Thus, it is not possible to disentangle in our analysis which specific action (if cash transfers, relief for utility bills, extended pension benefits, etc.) has been the one most likely to have reduced conflict. We nonetheless can ascertain that from the 24 analysed countries with COVID-19 welfare and labour policies, roughly 70% have implemented cash-transfers and 30% provided relief in paying utility bills. Also, the countries with a broader net of COVID-19 economic support, with more initiatives, are reducing the most the probability of experiencing conflicts and associated fatalities.

Our results also resonate with the earlier literature on food vulnerability, proxied by changes in local prices, and conflict (Brück & d'Errico, 2019; Jones, Mattiacci, & Braumoeller, 2017; Rezaeedaryakenari et al., 2020). The vast majority of food consumed in Africa (90%) comes from domestic producers (Clionad Raleigh, Choi, & Kniveton, 2015). Theoretically, one could argue that rises in local prices might benefit local producers. In reality, most producers in Africa are net consumers of food, which explains why increases in food prices can severely fuel conflicts. Thus, overall our findings highlight the importance of providing urgent welfare and labour assistance to curb conflicts.

The paper continues as follows. Section 2 provides an overview of the literature. We do not attempt to provide a discussion of the extensive literature on conflict, which can be found in detailed reviews (e.g. Blattman & Miguel, 2010; Collier & Hoeffler, 2014). Instead, we provide a summary of the conflict studies with relevance for COVID-19. Section 3 describes our data and instruments. Section 4 describes the econometric method used. Section 5 shows the results. Section 6 presents our conclusions.

## 2. LOCKDOWNS, PRICE VOLATILITY AND COVID-19 ASSISTANCE

For millions of people, the immediate concern is not the actual novel coronavirus itself, but surviving the economic hardship imposed by the lockdowns. Theoretically, there are at least three critical mechanisms by which lockdowns could fuel violent conflicts, despite the restrictions on population mobility. We describe these three important mechanisms (lockdowns, food vulnerability and welfare assistance) next.

### 2.1 Early social distancing measures and stricter lockdowns

Since the lockdowns, around the world some violent and non-violent crimes and conflicts have declined substantially.[2] However, in some countries, other conflicts have increased as the lockdowns intensified, such as riots and violence against civilians. Two key aspects could explain the rise in these conflicts. First, the existence of ongoing

---

[2] For instance, robbery and assault have plunged in Latin America, USA and European cities as the lockdowns limited population mobility and ease the job of the police in spotting and arresting suspects. Nonetheless domestic violence has risen, as well as cybercrimes (The Economist, 2020).

conflicts. Second, the way in which lockdowns have been enforced (with or without a safety net).

To address areas with pre-existing conflicts, the United Nations, on 23 March 2020, called for an immediate global ceasefire to allow medical personnel to reach the vulnerable population in these areas (UN News, 2020). The plea for a ceasefire has nonetheless been largely ignored. According to ACLED (2020) out of the 43 countries with at least 50 events of organised violence before lockdowns, only ten experienced unilateral ceasefire, another 31 countries experienced an increase in the rates of organised violence, such as Mexico, Iraq, Mozambique and Syria.[3]

Other conflicts emerged soon after lockdowns over food shortages such as in Lesotho, South Africa, Zimbabwe as citizens who suddenly lost their livelihoods desperately tried to get access to food parcels handed out by authorities (J. Burke, 2020). As governments face riots and revolts over food shortages and pleas for urgent assistance, there is a significant risk of using excessive force against civilians in the forms of military or police that could increase even further grievances and unrest. Even in countries without food riots or food lootings, governments risk using excessive force against civilians to enforce lockdowns.

*2.2 Food vulnerability*

Millions of people in Africa were already struggling to have enough to eat due to ongoing armed conflicts, extreme weather and long-historical institutional failures. However, this man-made imposed mobility restrictions to curb the COVID-19 pandemic add an extra layer of complexity. Perhaps one of the significant concerns of lockdowns is its effect on food vulnerability. Lockdowns have imposed tight mobility restrictions to farmers that have hampered efforts in delivering essential food and basic stables in at least 33 of Africa's 54 countries (Mutsaka, 2020). Although the pandemic has not disrupted the harvest per se, there are media reports of farmers in Africa with rotting crops as lorries have failed to arrive due to lockdown restrictions (Barrett, 2020;

---

[3] The abysmal response is perhaps not surprising. Ceasefires have slim chances of working in deeply entrenched conflicts, and in many instances, violence returns with a vengeance soon after (P. Burke, 2016). Although having a history of failed agreements surprisingly can lead to negotiating a ceasefire eventually, it is first required to have a record of failed attempts (Joshi & Quinn, 2015) that can only be built over time.

George, 2020). These lockdowns have also shut down many informal food markets where people earn their daily living, leaving large segments of the population without necessary provisions, and with real prospects of having not enough to eat. Moreover, school closures will also imply that nearly 370 million children worldwide risk missing out on school meals provided by the World Food Programme (WFM, 2020).

Major food supply chains have been a catalytic feature of many historical conflicts ranging from the French Revolution until the violent unrest that eventually led to the Arab Spring (Barrett, 2020). As such, there is an extensive literature detailing how sudden food insecurity leads directly or indirectly to violent riots and social unrest (Brück & d'Errico, 2019; Jones et al., 2017; Clionad Raleigh et al., 2015; Rezaeedaryakenari et al., 2020). According to this literature, there are at least three critical channels through which food vulnerability increases riots and violence against civilians.

First, at the individual level, food vulnerability deprives people from the most basic human right, enhances grievances and highlights differences in food entitlements, among those who can afford the luxury of food stuck for weeks and those who cannot even afford a meal a day (Hendrix & Brinkman, 2013; Jones et al., 2017). Survival instincts and grievances reduce the opportunity costs of engaging in violent riots against government, food looting and even joining rebel groups recruiting people in exchange for food and economic support during quarantines. Similar exchanges of food and "COVID-19 support packages" have been seen in Italy and Mexico with mafias and drug cartels, which are highly unlikely to be given without any form of expected reciprocity (Tondo, 2020). The literature has also reported such rebel and organised crime tactics in connection to food vulnerability and conflict in Africa (Humphreys & Weinstein, 2008). Rises in local food prices are a good proxy for food shortages and food vulnerability. Although theoretically, producers could benefit from an increase in prices, in the African context, most producers are net consumers of food, hence rises in local and international prices make producers worse off given the higher net cost of the food basket (Lee & Ndulo, 2011).[4]

---

[4] This negative effect is the case for most African states since they are neither major importers nor exporters (Clionad Raleigh et al., 2015). Similarly, an increase in local prices worsens food insecurity of consumers by reducing their ability to procure essential food to survive (Jones et al., 2017).

Second, at the rebels group level, food vulnerabilities also have a direct impact on the group ability to mobilise resources to support activities. Some rebel groups might have also lost substantial revenues from the sudden drop in prices of natural resources which they might have illegal access to such as oil. With such falls in profits, rebel groups have higher incentives to victimise ordinary citizens seeking resource appropriation, such as food. The areas with the largest share of cultivation are most susceptible to such rebel tactics, particularly during food shortages (Rezaeedaryakenari et al., 2020).

Third, at the national level, the government has a crucial role to play in dealing with food vulnerability and food-related conflicts. Governments might have different tolerance for food-related conflicts driven by ordinary citizens desperate for survival or if driven by rebel groups (Rezaeedaryakenari et al., 2020). Nonetheless, governments might use excessive violence against civilians to prevent further violent clashes and enforce strict lockdowns, depending on its ability to both provide adequate and urgent humanitarian support to struggling families during quarantines, and manage tactfully potential unrests.

*2.3 COVID-19 welfare and labour assistance*

Sudden lockdowns imposed without any safety net in place to help vulnerable populations risks pushing millions of people into extreme poverty and are likely to fuel conflicts. Developing countries are particularly constrained given the recent devaluation of many of their currencies, plummeting oil prices, and the collapse of major economic sectors. Despite this dark economic scenario, over 159 countries have implemented urgent welfare assistance and labour policies to deal with COVID-19 (Gentilini et al., 2020). The extent to which these packages manage to restrain significant increases in poverty and conflict will depend on their outreach. That is whether the extended COVID-19 welfare net can support households in difficult-to-reach rural areas, entrenched in conflicts.

From the vast literature on conflict, we know a great deal about how economic crises and shocks increase civil conflicts, riots and violence against civilians (Blattman & Miguel, 2010; Miguel, Satyanath, & Sergenti, 2004). Related literature offers mixed evidence on the extent to which foreign aid and foreign food aid can reduce the incidence of conflicts. Various studies have found that aid can reduce conflicts as it increases popular support for governments and increases the cost of opportunity of joining rebel and insurgent groups (E. Berman et al., 2011; de Ree & Nillesen, 2009; Nielsen, Findley, Davis, Candland, & Nielson, 2011). However, other studies have also found that (food) aid can increase both the incidence and the duration of civil conflicts (Nunn & Qian, 2014). Anti-poverty transfers such as community-driven programmes and food aid supplies have also been found to increase the intensity of conflicts (Crost et al., 2014) as insurgent groups sabotage these programmes to prevent weakening their ability to recruit future members.[5] A similar positive association has been found between increased conflict and rural employment programmes (Khanna & Zimmermann, 2014).

A small but growing strand of the literature has also studied the link between conditional cash transfers and conflict. The evidence is again somehow mixed. Some countries with deeply entrenched conflicts have ongoing conditional cash transfers without showing any direct link, such as the case of Mexico (Gutiérrez-Romero & Oviedo, 2018). Nonetheless, conditional cash transfers designed with the implicit aim of dismantling guerrilla groups have been found successful in reducing conflicts.[6] The literature suggests that these type of anti-poverty programmes can reduce the capacity of insurgents to recruit combatants from villages, increase electoral support for the incumbent government (Labonne, 2013), and increase the cost of opportunity of joining illegal activities in settings with long-entrenched civil conflicts (Pena et al., 2017). Nonetheless, it is unclear the extent to which countries with high rates of extreme poverty and exacerbated food vulnerability due to lockdowns will respond to the urgent and wide range of welfare and labour COVID-19 assistance packages. Many of the urgent welfare packages introduced are unconditional cash transfers that have shown to reduce food vulnerabilities and poverty in Africa as well as in other developing regions, but with a lesser known effect on conflict (Chakrabarti et al., 2020; Tiwari et al., 2016).

---

[5] There is also mixed evidence on whether community driven programmes can indeed reduce poverty as they can be used for clientelistic purposes and suffer from corruption (Gutiérrez-Romero, 2013).

[6] An example of such initiatives is the conditional cash transfer introduced in Colombia in 1999 in response to the major economic crises that affected Latin America (*Familias en Acción*). This conditional cash transfer reduced the probability of conflict and demobilised combatants, mainly children aged 10-17 (Pena et al., 2017). Similar evidence has been found in the Philippines (Crost et al., 2016).

Similarly, it is unclear whether governments in the developing world will have to rely on excessive use of force to guarantee lockdowns and curb potential violent unrests. We address these questions in the next sections.

3. DATA

3.1 *Data on conflict*

The data for all the dependent variables used on conflict come from the Armed Conflict Location and Event Data Project (ACLED). ACLED collects real-time data on all reported political violence and protests around the globe using a range of sources such as government reports, local media, humanitarian agencies, and research publications (C. Raleigh & Dowd, 2016). It has the main advantage of providing georeferenced data at the sub-country level by day and month within each year.[7]

In this paper we focus exclusively on four types of conflicts: riots, violence against civilians, food-related conflicts, and more specifically, food looting reported in ACLED from 1 January 2015 until 2 May 2020. Riots are defined by ACLED as a violent form of demonstration. Violence against civilians is defined as any armed or violent group attacking unarmed civilians who are not engaged in political violence (C. Raleigh & Dowd, 2016). Governments, rebels, militias and rioters can all be involved in these violent acts against civilians that can include attacks, abduction, forced disappearance and sexual violence. Food-related conflicts are not directly categorised in the publically available ACLED dataset. However, we identify these food-related conflicts based on the detailed description of each of the events reported in ACLED.

We analyse the ACLED's conflicts reported on a daily basis, that is without doing any aggregation on a monthly or yearly basis by country. This fine level of granularity as when the conflicts took place allows us to exploit the variation with which early social distancing measures, lockdowns and welfare/labour COVID-19 policy responses were implemented across countries.

---

[7] ACLED provides the exact number of conflicts, associated fatalities, location, exact date and actors involved across six broad types of conflict (which can be sub-categorised further). These six types of conflicts are: battles, explosions (e.g. suicide bombs, grenades), violence against civilians, protests, riots and strategic developments (e.g. non-violent actions on agreements, arrests, disrupted weapons use, etc).

3.2 *Dates of social distancing and lockdowns measures*

As COVID-19 spread around the globe, a wide range of social distancing measures and more strict lockdowns have been implemented. We obtain the exact date on which the first ever social distancing was implemented as well as the date of local lockdowns[8] from the publically available data on COVID-19 Government Response Tracker (OxCGRT), by Hale et al. (2020).[9] At the time of writing this paper, OxCGRT did not include data on social distancing measures for 13 African countries (Benin, Burundi, Central African Republic, Equatorial Guinea, Eritrea, Guinea, Guinea-Bissau, Ivory Coast, Liberia, Republic of Congo, Senegal, Somalia and Tongo).[10] For all these 13 countries, we took information on the exact date of early social distancing and lockdown from ACAPS (2020). From this database, we also took the period of the lockdown of Nigeria. Table A.1, in the Appendix, lists the dates of early social distancing and lockdowns for the countries we focus on in this paper.

---

[8] We obtain the exact date of lockdowns based on the date in which any of the eight reported social distancing measures took the highest ordinal value of 4, signalling the severity of lockdown.

[9] OxCGRT provides exact dates on when each of the social distancing measures were implemented across 149 countries, from January 2020 until 29 April 2020. This database contains the exact data on eight types of social distancing. These include: international travel restrictions, limitations on internal movement, closure of schools, closure of workplace, cancellations of public events, restrictions of large gatherings, stay at home requirements and restrictions on public transport. This information was collected from media, government reports and other publicly available sources. Another advantage of this dataset is that it provides an ordinal value of 1-4 to each of the eight social distancing implemented that helps to ascertain the level of their severity. The full methodology on how this dataset was collected and is being developed is available in the live report provided by (Hale et al., 2020).

[10] We do not have data on local prices for all these additional countries, but the dates of their lockdowns help doing the preliminary spatial analysis as well as the regression discontinuity plots presented in section 3.

3.3 *Constructing a monthly local index of prices at the market level*

To measure the link between food vulnerability and conflict, we use data from the Global Food Prices Database (WFP). This dataset reports monthly commodity prices at a sub-country level, across 985 local markets, in 23 African countries from the 1990s until May 2020 for which there is also information on conflicts in ACLED. We add information for Zimbabwe not included in WFP, from the USAID FEWS-NET dataset that also provides monthly local food prices. We focus our analysis on the 24 African countries, listed in Table A.1 that shows the countries for which we have data on local prices from 1 January 2015 until 2 May 2020.

The two sources of local prices used report a wide range of commodities, which are often not consistent across countries given the differences in diet and staple foods. Thus, we construct instead an index of monthly price of the most frequent commodity within each market.[11] This approach has also been used in the literature to overcome the variance in commodity baskets within and across countries (Clionad Raleigh et al., 2015). In our econometric analysis, we take January 2015 as the base for the index for each market, which allows us to assess to what extent the index of local prices has changed since then. For each conflict reported in ACLED we add the local price index of their closest food market within the same month, year and country where the conflict took place.

3.4 *Constructing an index of welfare and labour COVID-19 policy*

We construct an overall welfare and labour index based on these 12 different types of interventions implemented to deal with COVID-19, compiled by (Gentilini et al., 2020). By the period of our analysis, 1 May 2020, a total of 159 countries had implemented some sort of welfare and labour COVID-19 policy.[12] We use a simple additive

---

[11] That is for each market we construct a consumer price index as the sum of the total expenditure of most common items sold by multiplying price times quantity and adding them. The basket compared in each market is such that can be comparable over time. Then we divide the monthly consumer price index by the value of the index in the base year (January 2015).

[12] These can be grouped into three broad categories. The first one, social assistance interventions include: cash-based transfers, public works, in-kind/school feeding and utility/financial support. The second, social insurance policies include: paid

unweighted index to measure the whole range of various welfare and labour COVID-19 policy response.[13] In theory our index can take values from 0 (no intervention) up to 1 (a country that has taken all 12 types of interventions). In practice, the overall index ranges from 0 to slightly above 0.4 (that is, with five ongoing policies). Since Gentilini et al. (2020) do not include the exact date as when these interventions have been put in place, we take this information instead from Hale et al. (2020).[14]

Table A.2 in Appendix lists the welfare and labour policies implemented in each of the 24 African countries we focus on. From the 19 countries with an on-going COVID-19 welfare and labour policy, 12 have provided cash-transfers (among other policies); while the other seven have provided utility and financial support. Labour interventions are the least used thus far. Among the 24 countries analysed, only Egypt has adopted recent labour regulations.

---

leave/unemployment, health insurance support, pensions and disability benefits and social security contributions. The last one, labour market interventions: include wage subsidy, training, labour regulation and reduced work time subsidy.

[13] Various methods can be used to create composite indices such as additive, multiplicative and weighting some aspects with principal components analysis (Hale et al., 2020). We use the additive method as there are few interventions which might not merit using principal component analysis. We are not interested either in which policy explains the most variance in responses, rather to simply come with an index that measures the whole range of interventions in each country, which has the advantage of being simpler to interpret.

[14] To construct the index of welfare and labour COVID-19 response packages we prefer to use Gentilini et al. (2020) given the more extensive list of actions and programmes taken in each country 12 concrete actions over four categories of actions reported in Hale et al. (2020).

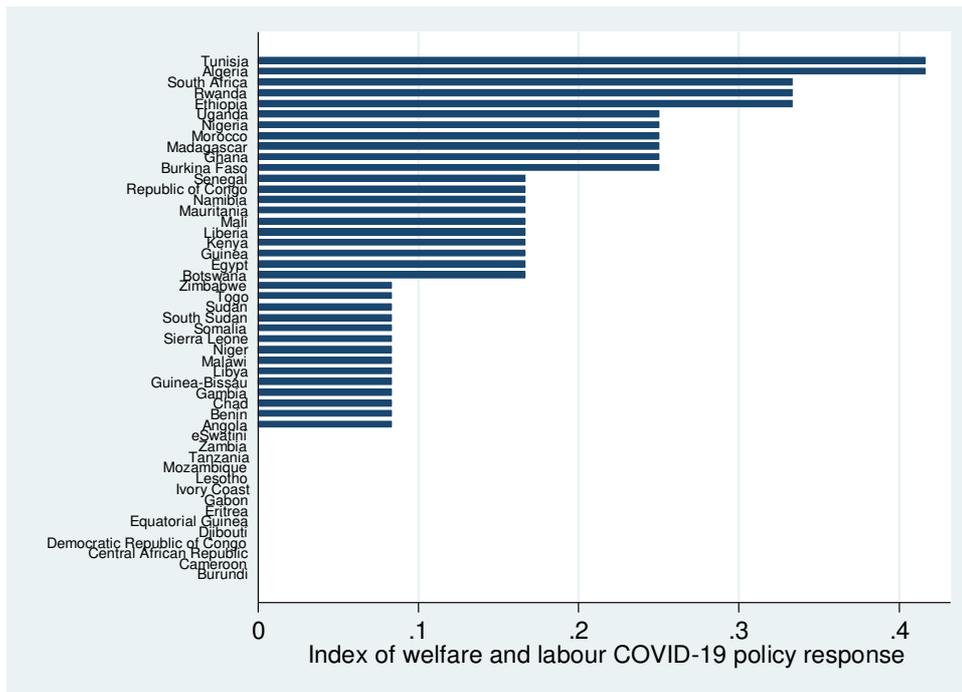

Figure 1: COVID-19 policy response index across Africa, as of 1 May 2020
Source: Own estimates using Gentilini et al. (2020).

3.5 *Other important controls at sub-country level*

Based on the extensive literature on conflict, we also include a wide range of control variables to mitigate potential confounding or unobserved characteristics. At the country-level, we include the ethnolinguistic fractionalisation index. At the district level, we use the monthly average of the stable nightlight luminosity from the DMSP-OLS Nighttime Light, from the USA Air Force Weather Agency. To avoid potential endogeneity issues, we use the monthly nightlight for the year 2015 only. We also use the log of the cultivated district, and size of the area (district) taken from the publicly available data from Rezaeedaryakenari et al. (2020). The remaining controls are drawn from the publicly available data from Manacorda and Tesei (2020) that allows us to construct data on georeferenced areas of on average about 55x55 km to each of the conflict events reported in ACLED. The variables used are the mobile phone coverage of 2G-3G, percentage of mountains, percentage of forests, the existence of petroleum fields, mines, diamond mines, electricity coverage, primary roads coverage, population and infant mortality rate. In Table A.3 we list the sources of each variable.[15] These variables help us to control for natural resources conflicts (N. Berman, Couttenier, Rohner, & Thoenig, 2017; Fenske & Zurimendi, 2017). Population size and mountains are also among the most relevant and statistically significant controls in the conflict literature (Collier & Hoeffler, 2014). Mobile phone coverage has been found crucial for political mobilisation and riots (Manacorda & Tesei, 2020). Similarly, the density of roads is important for the spatial distribution of conflict in Africa (Detges, 2016).

3.6 *Instrumental variables*

COVID-19 interventions have been highly dependent on political and economic contexts. Hence it would be hard to argue that social distancing, lockdowns and welfare/labour COVID-19 policy response have been exogenous or independent from existing conflicts within each country. For this reason, our econometric specification focuses on using instrumental variables. We use four instruments. We use the male mortality rate attributed to household and ambient air pollution per 100,000, based on standardised age, at the national level for the year 2016 and the percentage of diabetes prevalence among the adult population (aged 20 -79) at the national level over the years 2010-2019. Both instruments have been found in the medical literature as risk factors to COVID-19 (Fattorini & Regoli, 2020; Hussain et al., 2020), thus are likely to influence state's decision as to when to impose social distancing measures and the severity of lockdowns. We also include the IMF overall commodity monthly price index over the years 2015-2020 (including food, agriculture, fuel and non-fuel prices). This index is representative of the global market and is determined by the largest import market of a given commodity. This overall index helps to denote the severity of external fluctuations which might affect how countries respond to adopt different welfare and labour COVID-19 policies. The extent of the generosity of these packages is likely to depend on existing welfare structures and institutions, thus is likely shaped by colonial heritage (Nash & Patel, 2019). Hence, we also include a series of dummy variables denoting whether the country is a former British, French, Portuguese, German, Belgian

---

[15] We do not describe these variables in detail here as we refer the reader to the detailed description available in Manacorda and Tesei's article. We take this information for the latest year available in their series, year 2012.

or American Colonisation Society colony. Table A.3 lists the sources of these instruments.

3.7 *Description of conflicts*

We start by providing a broad description of the conflicts reported in ACLED for the entire African continent from 1 January 2015 until 2 May 2020. Figures 2, 3 and 4 show that soon after lockdowns the incidence of riots, violence against civilians and food-related conflicts increased when compared to the incidence of these conflicts to the period before lockdowns.

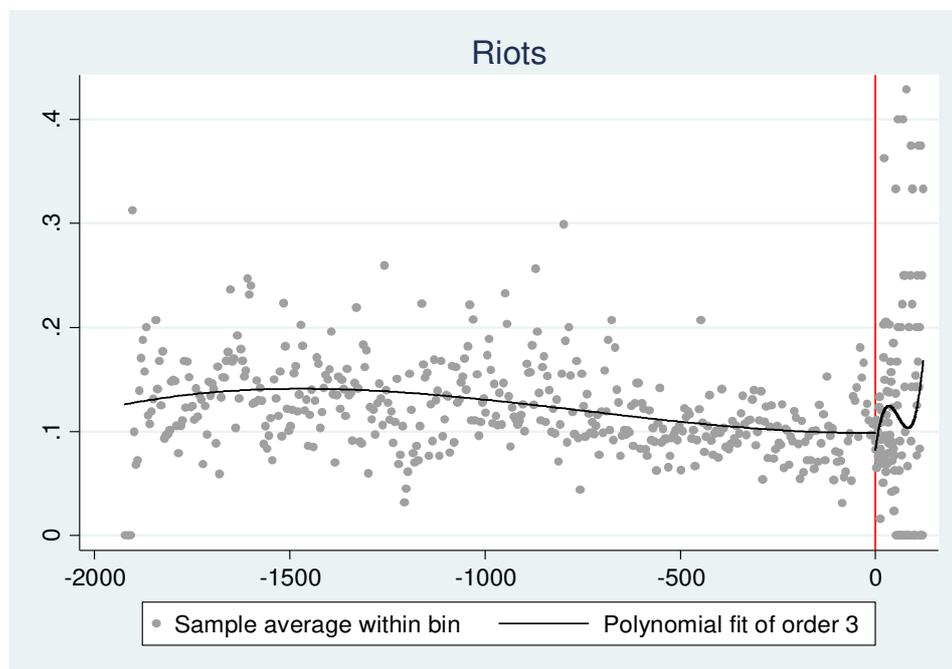

Figure 2: ACLED's riots before and after lockdown.

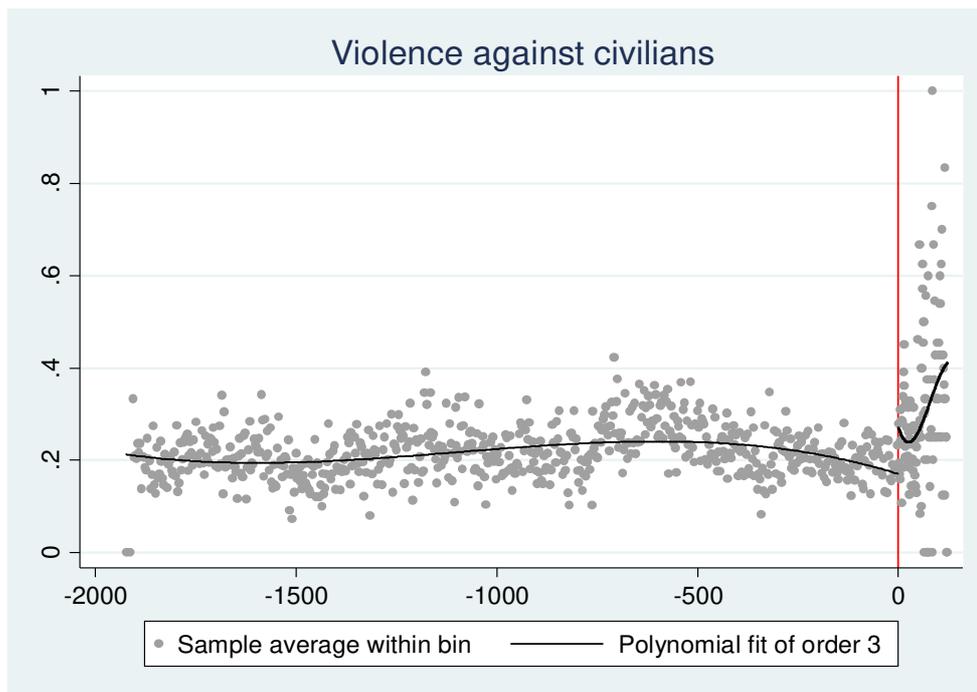

Figure 3: ACLED's violence against civilians before and after lockdown.

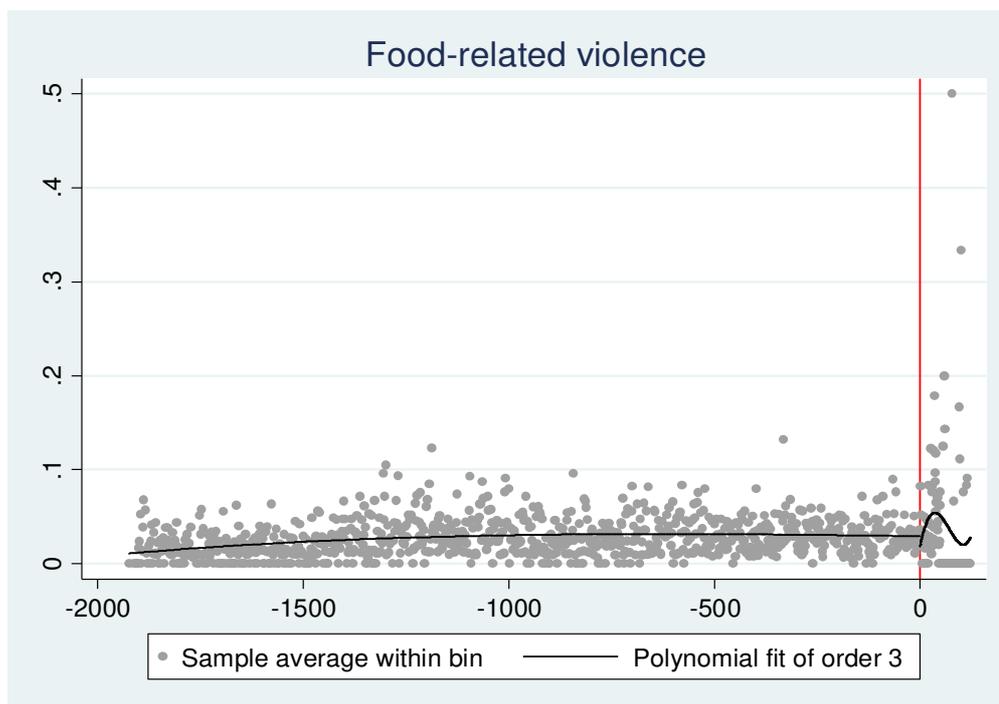

Figure 4: ACLED's food-related violence before and after lockdown.

Figure 5 illustrates the spatial distribution of riots, violence against civilians and food-related violence before (1 January 2015-before lockdown) and after lockdowns. These figures suggest that conflicts that erupted after lockdowns are more spatially concentrated in areas that had already ongoing conflicts. The maps on the right side of Figure 5 also shows that food-related conflicts are more concentrated in areas with a higher share of cultivated land (denoted by a darker colour). In a pre-COVID study, Rezaeedaryakenari et al. (2020) had noted the same spatial correlation. They suggested that the areas with more cultivation provide greater utility for forcible appropriation by rebels for the acquisition of food. When the aggregate food supply shrinks, as is likely the case after lockdowns, these geographical regions become a priority target.

Since we are concerned with the role of food volatility, the rest of our analysis focuses exclusively on the 24 countries for which we have data on local food prices. Table A.4 provides a summary description for these 24 countries from 1 January 2015 until 2 May 2020. In total there are 42,010 conflicts reported (including battles, explosions (e.g. suicide bombs, grenades), violence against civilians, protests, riots and strategic developments. About a third of these events (28%) were violence against civilians and nearly a quarter (13%) was riots, with a minority of food-related conflicts and food looting (2%). The state has been involved as an actor in nearly 32% of *all* reported ACLED conflict cases.

The total and average of the fatalities per event are also reported in Table A.4. In total there were 169,454 fatalities associated with *any* conflict reported in ACLED, from 1 January 2015 until 2 May 2020. There were 4,552 fatalities associated with riots, 50,506 fatalities associated with violence against civilians and 6,888 fatalities associated with *any* food-related conflicts (including food looting), with 4,344 fatalities due to food looting.

Figure 6 illustrates the potential link between violence against civilians, local food prices and the IMF global commodity index. We focus only on the 24 countries for which we have local food prices. Only for Figure 6 we aggregate the data at monthly level for each country. We also standardise each of the three depicted variables such that their monthly average is divided by the maximum value of each variable for the entire series. Thus, the y-axis shows how much the monthly series fluctuates from the highest level achieved within each country.

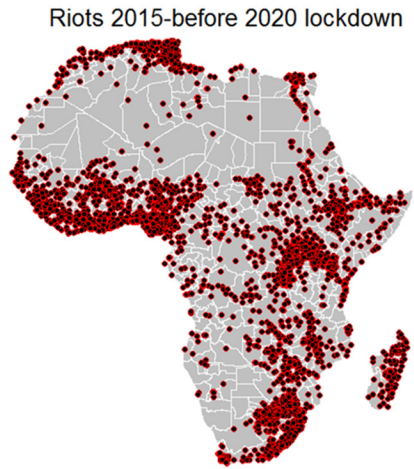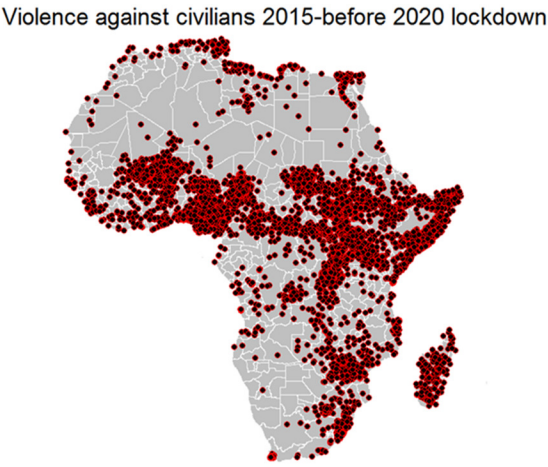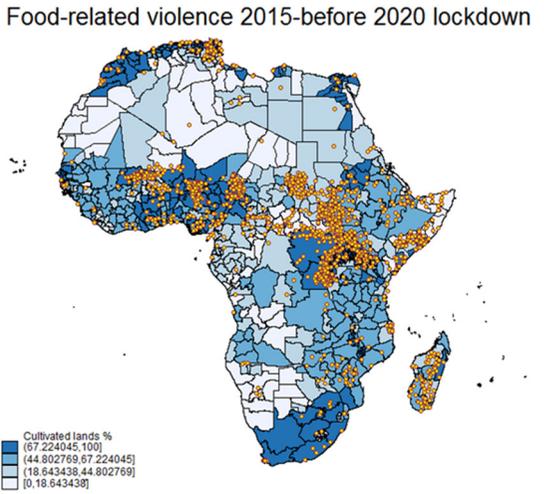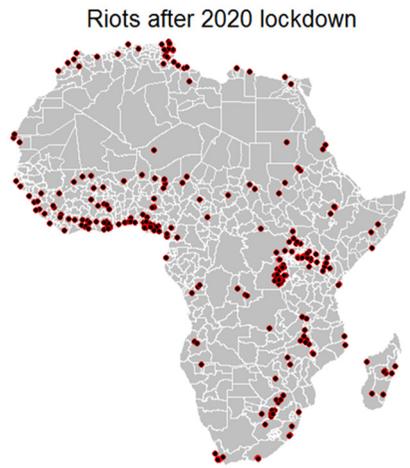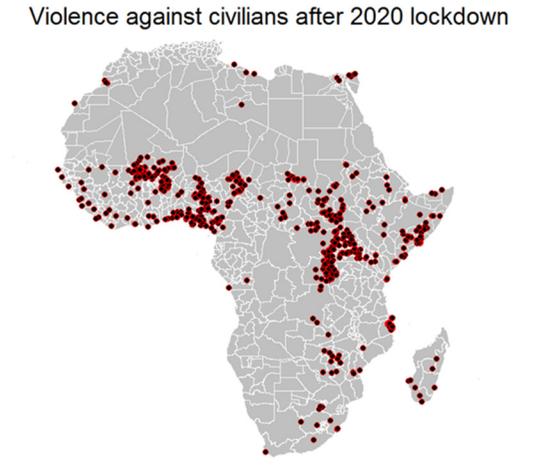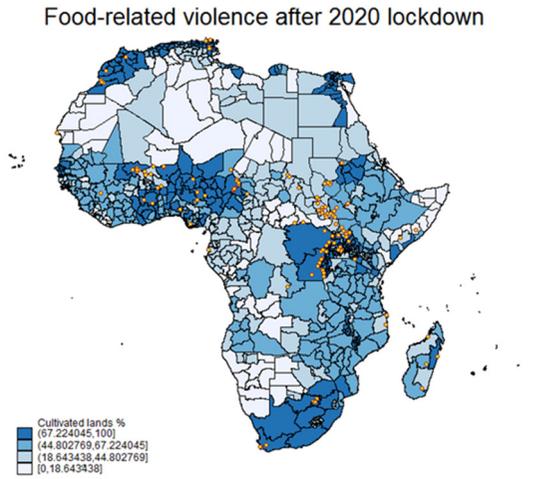

Figure 5. Conflict and lockdowns

For some countries, there is a particularly strong correlation between local food prices and conflicts such as Ethiopia, Nigeria and Rwanda. However, there are many exceptions where the local prices have increased, whereas violence against civilians has not. That is the case of Burkina Faso, Malawi and Namibia. This evidence might suggest that albeit rises in food prices might have contributed to some conflicts, but the welfare and labour COVID-19 interventions could have dampened some of the violence against civilians. We analyse these issues next.

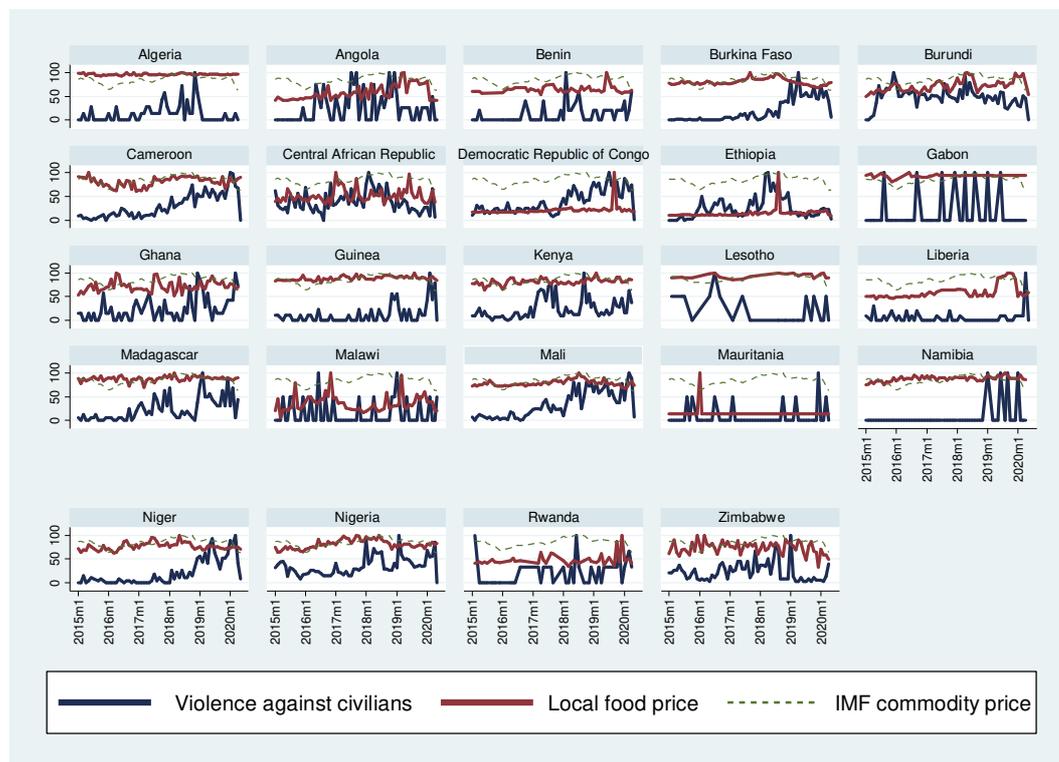

Figure 6: Monthly prices and ACLED's violence against civilians.

4. ESTIMATION FRAMEWORK

We use two econometric specifications to estimate the impact of social distancing measures, price volatility and welfare and labour COVID-19 policies on conflict. First, we use a panel random effects (RE) model, as shown in equation (1). The RE model has two main advantages. First, this specification can simultaneously model both time-variant and time-invariant effects (Bell & Jones, 2015). Second, the RE specification can deal with hierarchical data (in our case having repeated observations in sub-country level, cells, nested within countries, the higher-level fixed units), the reason why this specification is also known as the multilevel, hierarchical or mixed model.

$$conflict_{jit} = \alpha + \delta_1 S_{it} + \delta_2 \log local\ price_{jit} + \delta_3 X_{ji} + \delta_4 C_i + (\eta_{ji} + \varepsilon_{jit}) \qquad (1)$$

We focus on the incidence of four types of conflicts: riots, violence against civilians, food-related conflict incidents and food looting in the cell $j$ (with reported latitude and longitude in ACLED) located in country $i$ in day, month and year $t$ ($conflict_{jit}$). Our dependent variable is binary for each of the four types of conflicts analysed. $S_{it}$ is a vector that includes the three COVID-19 interventions we focus on: the first social distancing measure implemented[16], local lockdown measure and the welfare and labour COVID-19 policy index in country $i$ implemented at day, month, year $t$. The first social distancing measure refers to the date in which this was implemented. Lockdown takes the value of 0 or 1 depending on if the conflict occurred before or after the respective lockdown. The monthly local price index (measured in log) at cell $j$ in country $i$ ranges from January 2015 until 2 May 2020. $X_{ji}$ is a vector that captures our controls at the cell $j$ located in country $i$ and includes: the percentage of mountains, forests, whether the cell has petroleum fields, mines, diamond mines, size of the area (district level). In addition, vector $X$ includes some key variables lagged in time to mitigate potential endogeneity issues. These lagged variables are the stable nightlight (measured in log for the year 2015), the percentage of mobile phone coverage in 2G-3G, the percentage of electricity coverage, primary roads coverage, population, infant mortality, percentage of land cultivated. Vector $C$ includes the ethnolinguistic fractionalisation index, at country level $i$. ($\eta_{ji} + \varepsilon_{jit}$) denotes the time-invariant and time-variant error term. The results of the RE specifications are shown in Table 1, columns 1-4.

The RE estimates will be unbiased if there are no strong sources of endogeneity such as omitted variable due to unobserved heterogeneity. However, we suspect that the RE specifications are biased, given the unlikely exogenous characteristics of the three COVID-19 interventions we focused on. We therefore add to our RE specification IV-2SLS estimates to address this potential endogeneity. We instrument our three likely endogenous variables: the date of the first social distancing measure, whether in lockdown and the welfare and labour COVID-19 policy index denoted by $S_{it}$. Our instruments, denoted by vector $Z_{it}$, are: male mortality rate attributed to household and ambient air pollution per 100,000 (lagged for year 2016), diabetes prevalence (% of population ages 20 to 79, years 2010-2019), IMF all commodity price index (years 2015-2020), whether the country is a former British, French, Portuguese, German, Belgian or American Colonisation Society colony. The first-stage relationship between our three endogenous variables, $S_{it}$, and our instruments $Z_{it}$ are shown in equation (2).

---

[16] This index takes the value of 0 before any policy included in the index was implemented, and takes the value of the constructed index after the first welfare/labour COVID- response policy was implemented according to Hale et al. (2020).

$$S_{it} = \gamma + \mu_1 Z_{it} + \mu_2 log\ local\ price_{jit} + \mu_3 X_{ji} + \mu_4 C_i + v_{jit} \qquad (2)$$

The second-stage equation estimates the impact of the instrumented $\hat{S}$ COVID-19 responses on the incidence of conflict, as denoted by equation (3). The ($\xi_{ji} + \varphi_{jit}$) denotes the time-invariant and time-variant error term. We implement this IV regression using panel random effects.

$$conflict_{jit} = \kappa + \beta_1 \hat{S}_{it} + \beta_2 log\ local\ price_{jit} + \beta_3 X_{ji} + \beta_4 C_i + (\xi_{ji} + \varphi_{jit}) \qquad (3)$$

The results of the second-stage IV-2SLS regression are reported in Table (1), in columns 5-8. At the bottom of the table, we report the Sargan-Hanssen overidentification tests. The null hypothesis of this test is that the over-identifying restrictions are valid. We also present the Hausman endogeneity test. The first-stage regression is shown in Table A.5. In sum, all our instruments are strongly correlated to the endogenous variables, satisfy the overidentification tests. There is evidence that the COVID-19 measures of lockdowns and welfare assistance are endogenous, hence implemented in response to conflicts, in particular in column 6 and 7 (violence against civilians and food-related incidents). Therefore these second-stage IV 2SLS regressions are our preferred specifications.

# 5. RESULTS

## 5.1 *Riots, violence against civilians and food-related conflict*

Early social distancing measures are not statistically significant with the incidence of the conflicts analysed, riots, violence against civilians food-related conflicts and food looting. That is the case in the random specifications RE with and without using instrumental variables (Table 1, columns 1-8). The non-significant effect is unsurprising since many of these early measures did not impose any mobility restrictions on the population but mostly focused on having some travel restrictions from abroad. The stricter lockdown measures yield different results. If focused on the IV-2SLS results, Table 1, columns 5-8, show that the probability of experiencing riots, violence against civilians, food-related conflicts and food looting did increase after lockdowns, as our earlier figures 2, 3 and 4 had shown.

Table 1 also shows that contemporaneous changes in prices are positively and statistically associated with violence against civilians (but not to riots, food-related conflicts or food looting). Specifically, a 10% increase in the value of the local price index is associated with a rise of 0.71 percentage point increase in violence against civilians. The same results are obtained when using the RE specifications with or without instrumenting. Among other variables prominently cited in the literature, we can conclude that riots are more likely to occur in more urbanised settings as they have higher levels of stable nightlight, mobile phone, electricity coverage and population. In contrast, violence against civilians seems to be concentrated in less urbanised settings as they have lower levels of stable nightlight, less electricity coverage, primary roads, but more cultivated land and mines.

Food-related incidents and food looting are more likely to occur in areas with a greater density of cultivated land, as Figure 5 suggested. However, the volatility of local prices is not associated with these food-related conflicts. These areas seem to be less urbanised as they have less density of primary roads, electricity.

There is also strong evidence from the IV-2SLS specifications that the welfare and labour COVID-19 policy index has reduced the probability of riots, violence against civilians and food-related conflicts, including food looting. For instance, Figures 7, 8 and 9 show the marginal effect of the probability of experiencing riots, violence against civilians and food-related conflicts with the values of the welfare and labour COVID-19 policy index. These marginal effects depict the IV-2SLS specifications shown in Table 1, columns 5-7. The effect of the index is negative and linearly associated with the probability of experiencing riots. Specifically, a 0.1 unit increase in the welfare/labour COVID-19 policy index, the likelihood of experiencing these conflicts declines by nearly 0.2 percentage points.

Table 1. *COVID-19 interventions, local prices and conflict*

|  | (1) | (2) | (3) | (4) | (5) | (6) | (7) | (8) |
|---|---|---|---|---|---|---|---|---|
|  | Panel Random Effects (RE) | | | | Panel RE IV specifications | | | |
|  | Riots | Violence against civilians | Food-related incidents | Food looting | Riots | Violence against civilians | Food-related incidents | Food looting |
| First social distancing implemented | 0.002 | -0.001 | -0.000 | -0.000 | 0.002 | -0.003 | -0.001 | -0.000 |
|  | (0.001) | (0.001) | (0.000) | (0.000) | (0.004) | (0.003) | (0.001) | (0.001) |
| Strict lockdown | 0.020* | 0.078*** | 0.021*** | 0.009** | 0.154*** | 0.110* | 0.127*** | 0.050*** |
|  | (0.010) | (0.014) | (0.005) | (0.004) | (0.045) | (0.060) | (0.022) | (0.017) |
| Index of welfare and labour COVID19 response | -0.048 | 0.066 | -0.048** | -0.030* | -0.666** | -2.124*** | -0.886*** | -0.394*** |
|  | (0.048) | (0.063) | (0.023) | (0.018) | (0.282) | (0.378) | (0.138) | (0.108) |
| Log index local market price | -0.004 | 0.073*** | -0.001 | -0.000 | -0.004 | 0.071*** | -0.000 | -0.000 |
|  | (0.005) | (0.006) | (0.002) | (0.002) | (0.005) | (0.006) | (0.002) | (0.002) |
| Log stable nightlight (year 2015) | 0.012*** | -0.051*** | 0.002 | 0.002 | 0.013*** | -0.049*** | 0.003* | 0.003* |
|  | (0.004) | (0.005) | (0.002) | (0.001) | (0.004) | (0.005) | (0.002) | (0.001) |
| Log mobile phone coverage 2G-3G | 0.027*** | -0.005 | -0.003*** | -0.003*** | 0.027*** | -0.005 | -0.003** | -0.003*** |
|  | (0.002) | (0.003) | (0.001) | (0.001) | (0.002) | (0.003) | (0.001) | (0.001) |
| % Mountains | -0.056*** | 0.035*** | -0.010*** | -0.009*** | -0.057*** | 0.029*** | -0.011*** | -0.009*** |
|  | (0.008) | (0.010) | (0.004) | (0.003) | (0.008) | (0.010) | (0.004) | (0.003) |
| % Forests | 0.035*** | -0.039*** | -0.044*** | -0.031*** | 0.035*** | -0.043*** | -0.044*** | -0.031*** |
|  | (0.010) | (0.013) | (0.005) | (0.004) | (0.010) | (0.014) | (0.005) | (0.004) |
| Petroleum fields | 0.023** | 0.062*** | -0.007 | -0.004 | 0.021** | 0.055*** | -0.010** | -0.005 |
|  | (0.009) | (0.012) | (0.004) | (0.004) | (0.009) | (0.013) | (0.005) | (0.004) |
| Mines | -0.010*** | 0.021*** | -0.002 | -0.001 | -0.010*** | 0.020*** | -0.003* | -0.001 |
|  | (0.003) | (0.004) | (0.001) | (0.001) | (0.003) | (0.004) | (0.001) | (0.001) |
| Diamond mines | 0.003 | -0.017** | -0.001 | -0.001 | 0.004 | -0.015** | -0.000 | -0.000 |
|  | (0.005) | (0.007) | (0.003) | (0.002) | (0.005) | (0.007) | (0.003) | (0.002) |
| Size of area | 0.000 | 0.000** | 0.000*** | 0.000*** | 0.000 | 0.000** | 0.000*** | 0.000*** |
|  | (0.000) | (0.000) | (0.000) | (0.000) | (0.000) | (0.000) | (0.000) | (0.000) |
| Electricity | 0.045*** | -0.061*** | -0.009*** | -0.006*** | 0.046*** | -0.063*** | -0.009*** | -0.006*** |
|  | (0.006) | (0.007) | (0.003) | (0.002) | (0.006) | (0.007) | (0.003) | (0.002) |
| Primary roads | -0.011*** | -0.016*** | -0.003** | -0.003*** | -0.011*** | -0.019*** | -0.003** | -0.003*** |
|  | (0.003) | (0.004) | (0.001) | (0.001) | (0.003) | (0.004) | (0.001) | (0.001) |
| Log population | 0.010*** | -0.003 | -0.002* | -0.001 | 0.010*** | -0.001 | -0.001 | -0.001 |
|  | (0.002) | (0.003) | (0.001) | (0.001) | (0.002) | (0.003) | (0.001) | (0.001) |
| Log infant mortality rate | 0.000 | -0.140*** | 0.035*** | 0.026*** | 0.002 | -0.147*** | 0.039*** | 0.029*** |
|  | (0.011) | (0.014) | (0.005) | (0.004) | (0.011) | (0.015) | (0.005) | (0.004) |
| Log cultivated | -0.001 | 0.025*** | 0.014*** | 0.010*** | -0.001 | 0.027*** | 0.015*** | 0.010*** |
|  | (0.004) | (0.005) | (0.002) | (0.001) | (0.004) | (0.005) | (0.002) | (0.001) |
| Ethnolinguistic fractionalisation index | 0.107 | -0.125 | -0.014 | -0.017 | 0.110 | -0.186 | -0.030 | -0.027 |
|  | (0.137) | (0.100) | (0.024) | (0.016) | (0.242) | (0.184) | (0.048) | (0.036) |
| Constant | -37.788 | 22.619 | 6.272 | 5.268 | -41.348 | 57.267 | 13.611 | 10.661 |
|  | (29.795) | (21.332) | (4.964) | (3.256) | (94.867) | (71.291) | (18.369) | (13.672) |
| Observations | 42,010 | 42,010 | 42,010 | 42,010 | 42,010 | 42,010 | 42,010 | 42,010 |
| Number of countries | 24 | 24 | 24 | 24 | 24 | 24 | 24 | 24 |
| Test of overidentification restrictions: | | | | | | | | |
| Sargan-Hanssen statistics Chi-sq(1) | | | | | 2.134 | 9.463 | 4.772 | 3.615 |
| P-value | | | | | 0.907 | 0.149 | 0.573 | 0.730 |
| Hausman test | | | | | | | | |
| Chi2 | | | | | 11.350 | 167.050 | 46.530 | 21.680 |
| Prob>chi2 | | | | | 0.838 | 0.000 | 0.000 | 0.198 |

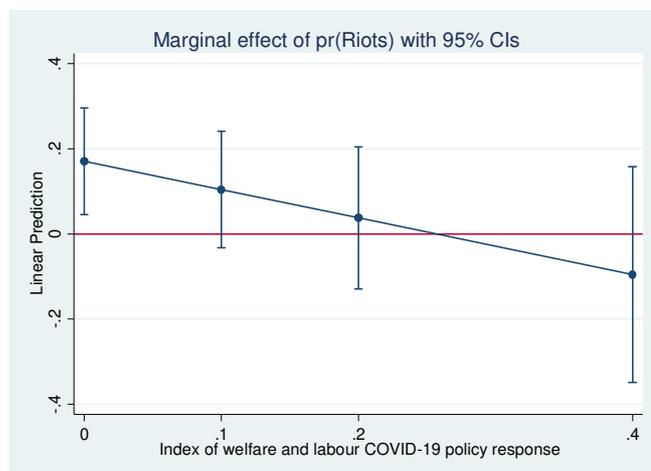

Figure 7: Riots and the welfare/labour COVID-19 policy index

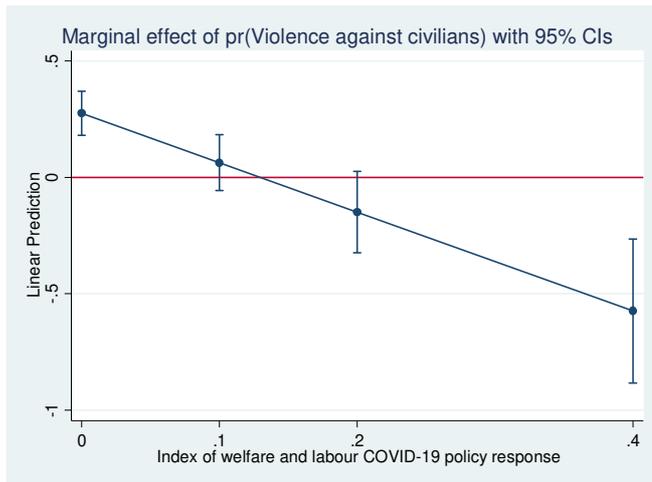

Figure 8: Violence against civilians and the welfare/labour COVID-19 policy index

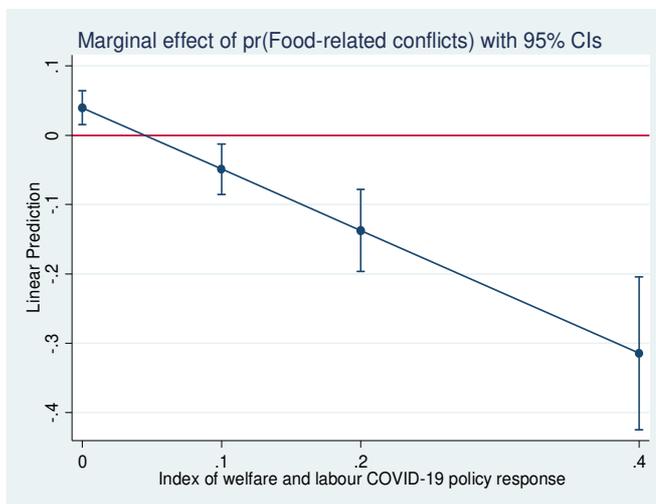

Figure 9: Food-related violence and the welfare/labour COVID-19 policy index

## 5.2 *Fatalities*

We next explore the total number of fatalities, as our new dependent variables to assess the magnitude of the conflicts analysed thus far. We analyse the number of fatalities reported in ACLED from 1 January 2015 until 2 May 2020 associated with *any* conflict. We also focus on the number of fatalities exclusively related to the conflicts of our interest: riots, violence against civilians and food-related conflicts (including food looting). As before, we use two specifications: panel random effects (RE) and panel random effects with IV-2SLS. Table 2 reports the results. As before at the bottom of the table, we report the Sargan-Hanssen overidentification test and the Hausman endogeneity tests. The first-stage regression results are reported in Table A.6. These first-stage regressions, along with the overidentification tests, suggest the instruments are valid. Again, we find evidence of endogeneity, particularly for all ACLED fatalities and fatalities due to violence against civilians (Table 2, columns 5 and 7).

The IV-2SLS specifications show that early social distancing measures have no increased association with fatalities (Table 2, columns 5-8). However, the number of fatalities increased substantially after lockdowns for all ACLED fatalities (columns 5), and fatalities associated with violence against civilians (column 7). There is no evidence of increased fatalities associated with food-related conflict. For this type of conflict, we added any fatalities associated with food looting as well.

There is evidence that countries with a higher welfare and labour COVID-19 policy index experienced lower levels of overall ACLED's fatalities as well as a lower level of fatalities due to violence against civilians (Table 2, columns 5 and 7). Figure 10 shows these marginal effects. For instance, the number of total fatalities, decrease by nearly ten casualties when comparing a country with no welfare and labour COVID-19 policy response versus one that has an index of 0.4.

As mentioned earlier (Table 1) higher local prices are not associated with a higher probability of experiencing food-related conflicts. However, Table 2, reveals that increases in local prices are associated with a higher number of fatalities due to food-related conflicts.

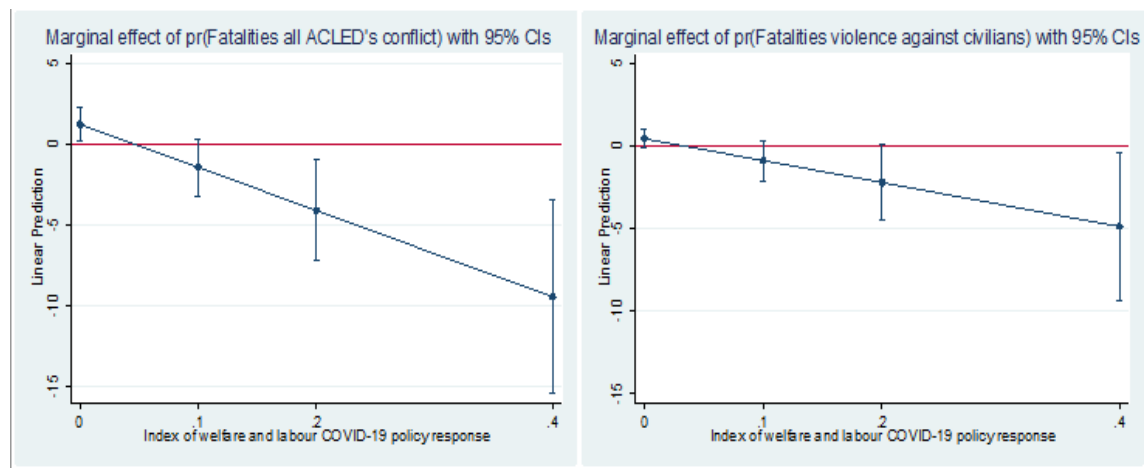

Figure 10: Overall fatalities and fatalities due to violence against.

Table 2. *COVID-19 interventions, local prices and fatalities*

| | (1) | (2) | (3) | (4) | (5) | (6) | (7) | (8) |
|---|---|---|---|---|---|---|---|---|
| | Panel Random Effects (RE) | | | | Panel RE IV specifications | | | |
| Fatalaties of: | Any ACLED conflict | Riots | Violence against civilians | Food-related conflict | Any ACLED conflict | Riots | Violence against civilians | Food-related conflict |
| First social distancing implemented | -0.024*** | 0.001 | -0.014*** | -0.002*** | -0.033 | 0.000 | -0.007 | -0.002*** |
| | (0.007) | (0.001) | (0.002) | (0.000) | (0.035) | (0.002) | (0.017) | (0.001) |
| Strict lockdown | 0.310 | 0.017 | -0.111 | 0.021 | 5.997*** | 0.192 | 2.396*** | 0.061 |
| | (0.277) | (0.029) | (0.208) | (0.035) | (1.206) | (0.129) | (0.911) | (0.167) |
| Index of welfare and labour COVID19 response | -2.057 | -0.052 | -0.864 | 0.102 | -26.693*** | -0.345 | -13.333** | -0.809 |
| | (1.266) | (0.134) | (0.948) | (0.160) | (7.537) | (0.804) | (5.692) | (1.048) |
| Log index local market price | -0.240** | -0.018 | 0.051 | 0.033*** | -0.240* | -0.019 | -0.020 | 0.035** |
| | (0.120) | (0.013) | (0.070) | (0.012) | (0.125) | (0.013) | (0.093) | (0.014) |
| Log stable nightlight (year 2015) | -0.037 | 0.010 | 0.103 | 0.001 | -0.001 | 0.010 | 0.145** | 0.011 |
| | (0.095) | (0.010) | (0.065) | (0.011) | (0.100) | (0.010) | (0.074) | (0.012) |
| Log mobile phone coverage 2G-3G | -0.692*** | 0.017*** | -0.105*** | -0.043*** | -0.708*** | 0.018*** | -0.085* | -0.040*** |
| | (0.059) | (0.006) | (0.038) | (0.006) | (0.061) | (0.006) | (0.045) | (0.007) |
| % Mountains | 0.113 | -0.053*** | -0.149 | -0.016 | 0.077 | -0.055*** | -0.155 | 0.001 |
| | (0.197) | (0.021) | (0.123) | (0.021) | (0.205) | (0.021) | (0.153) | (0.024) |
| % Forests | -2.189*** | -0.010 | -0.433*** | -0.157*** | -2.174*** | -0.016 | -0.754*** | -0.166*** |
| | (0.258) | (0.027) | (0.161) | (0.027) | (0.272) | (0.028) | (0.202) | (0.031) |
| Petroleum fields | -1.674*** | -0.045* | -0.424** | -0.042 | -1.809*** | -0.045* | -0.698*** | -0.063** |
| | (0.244) | (0.026) | (0.169) | (0.028) | (0.251) | (0.026) | (0.187) | (0.030) |
| Mines | 0.098 | -0.015* | 0.018 | -0.013 | 0.094 | -0.015* | 0.113* | -0.010 |
| | (0.078) | (0.008) | (0.057) | (0.010) | (0.079) | (0.008) | (0.059) | (0.010) |
| Diamond mines | 0.312** | -0.004 | 0.243*** | 0.009 | 0.360** | -0.003 | 0.289*** | 0.015 |
| | (0.137) | (0.014) | (0.101) | (0.017) | (0.141) | (0.015) | (0.105) | (0.017) |
| Size of area | 0.000*** | 0.000 | 0.000*** | 0.000* | 0.000*** | 0.000 | 0.000*** | 0.000** |
| | (0.000) | (0.000) | (0.000) | (0.000) | (0.000) | (0.000) | (0.000) | (0.000) |
| Electricity | 0.012 | 0.067*** | -0.018 | -0.040** | 0.072 | 0.066*** | 0.014 | -0.036** |
| | (0.142) | (0.015) | (0.093) | (0.016) | (0.147) | (0.015) | (0.110) | (0.017) |
| Primary roads | 0.062 | -0.021*** | -0.043 | -0.011** | 0.063 | -0.018** | -0.034 | -0.016** |
| | (0.069) | (0.007) | (0.033) | (0.006) | (0.074) | (0.008) | (0.055) | (0.007) |
| Log population | -0.433*** | 0.005 | -0.175*** | 0.000 | -0.444*** | 0.004 | -0.265*** | -0.004 |
| | (0.051) | (0.005) | (0.035) | (0.006) | (0.053) | (0.006) | (0.039) | (0.006) |
| Log infant mortality rate | 1.139*** | 0.040* | 0.343*** | 0.038** | 1.294*** | 0.046 | -0.087 | 0.071*** |
| | (0.238) | (0.024) | (0.092) | (0.015) | (0.282) | (0.028) | (0.205) | (0.026) |
| Log cultivated | 0.351*** | 0.002 | 0.396*** | 0.034*** | 0.321*** | 0.001 | 0.211*** | 0.029** |
| | (0.099) | (0.010) | (0.064) | (0.011) | (0.103) | (0.011) | (0.077) | (0.012) |
| Ethnolinguistic fractionalisation index | 0.321 | -0.013 | -0.113 | -0.093** | 0.137 | -0.024 | 0.105 | -0.143** |
| | (0.723) | (0.068) | (0.218) | (0.036) | (2.036) | (0.118) | (1.058) | (0.065) |
| Constant | 538.073*** | -15.911 | 302.793*** | 39.856*** | 725.467 | -10.325 | 153.514 | 53.191*** |
| | (143.598) | (13.415) | (39.689) | (6.589) | (759.140) | (40.408) | (376.244) | (18.644) |
| Observations | 42,010 | 42,010 | 42,010 | 42,010 | 42,010 | 42,010 | 42,010 | 42,010 |
| Number of countries | 24 | 24 | 24 | 24 | 24 | 24 | 24 | 24 |
| Test of overidentification restrictions: | | | | | | | | |
| Sargan-Hanssen statistics Chi-sq(1) | | | | | 0.499 | 3.081 | 2.140 | 4.716 |
| P-value | | | | | 0.998 | 0.799 | 0.906 | 0.581 |
| Hausman test | | | | | | | | |
| Chi2 | | | | | 40.080 | 11.960 | 104.890 | 18.850 |
| Prob>chi2 | | | | | 0.001 | 0.803 | 0.000 | 0.337 |

5.3 *The state as an actor in riots, violence against civilians and food-related conflicts*

To conclude our analysis, we focus on the conflicts in which the state has been directly involved as an actor (either instigating or responding to contain violence) and distinguish between riots, violence against civilians and food related conflicts. We identify whether the state was involved as an actor whether in its capacity as the military, the police, the government or government's guards. We obtain this information from the detailed notes revealed in ACLED's database.

As before we present two specifications, panel random effects (RE) and panel random effects with IV-2SLS. Table 3 presents both specifications, and Table A.7 shows the first-stage IV-2SLS specifications.

Table 3. *COVID-19 interventions, local prices and the state as perpetrator of violence*

| State (military, policy, gard or government) involved as actor in: | (1) Any ACLED conflict | (2) Riots | (3) Violence against | (4) Food-related conflict | (5) Any ACLED conflict | (6) Riots | (7) Violence against | (8) Food-related conflict |
|---|---|---|---|---|---|---|---|---|
| | **Panel RE IV specifications** | | | | **Panel RE IV specifications** | | | |
| First social distancing implemented | -0.002*** | 0.001 | -0.000 | -0.000 | -0.001 | 0.001 | 0.000 | -0.000 |
| | (0.001) | (0.001) | (0.000) | (0.000) | (0.002) | (0.002) | (0.001) | (0.000) |
| Strict lockdown | 0.071*** | 0.015** | 0.064*** | 0.006** | -0.300*** | 0.001 | -0.086*** | 0.018* |
| | (0.015) | (0.007) | (0.007) | (0.002) | (0.065) | (0.031) | (0.031) | (0.011) |
| Index of welfare and labour COVID19 response | 0.111 | -0.023 | 0.057* | -0.002 | 2.563*** | 0.104 | 0.486** | -0.126* |
| | (0.068) | (0.031) | (0.033) | (0.011) | (0.407) | (0.184) | (0.195) | (0.066) |
| Log index local market price | -0.016** | -0.003 | 0.010*** | 0.001 | -0.017*** | -0.003 | 0.009*** | 0.001 |
| | (0.007) | (0.003) | (0.003) | (0.001) | (0.007) | (0.003) | (0.003) | (0.001) |
| Log stable nightlight (year 2015) | 0.030*** | 0.017*** | -0.001 | -0.001 | 0.025*** | 0.017*** | -0.002 | 0.000 |
| | (0.005) | (0.002) | (0.003) | (0.001) | (0.005) | (0.002) | (0.003) | (0.001) |
| Log mobile phone coverage 2G-3G | -0.049*** | 0.011*** | 0.002 | -0.000 | -0.049*** | 0.011*** | 0.002 | -0.000 |
| | (0.003) | (0.001) | (0.002) | (0.001) | (0.003) | (0.001) | (0.002) | (0.001) |
| % Mountains | 0.047*** | -0.028*** | 0.035*** | -0.000 | 0.044*** | -0.028*** | 0.033*** | -0.000 |
| | (0.011) | (0.005) | (0.005) | (0.002) | (0.011) | (0.005) | (0.005) | (0.002) |
| % Forests | -0.059*** | -0.004 | 0.034*** | -0.008*** | -0.064*** | -0.005 | 0.033*** | -0.008*** |
| | (0.014) | (0.007) | (0.007) | (0.002) | (0.015) | (0.007) | (0.007) | (0.002) |
| Petroleum fields | -0.107*** | -0.012** | 0.003 | -0.003 | -0.098*** | -0.012* | 0.005 | -0.004 |
| | (0.013) | (0.006) | (0.006) | (0.002) | (0.014) | (0.006) | (0.007) | (0.002) |
| Mines | -0.007* | -0.003* | -0.004* | -0.000 | -0.005 | -0.003* | -0.003* | -0.000 |
| | (0.004) | (0.002) | (0.002) | (0.001) | (0.004) | (0.002) | (0.002) | (0.001) |
| Diamond mines | 0.019** | -0.000 | 0.003 | 0.001 | 0.014* | -0.001 | 0.002 | 0.001 |
| | (0.007) | (0.003) | (0.004) | (0.001) | (0.008) | (0.003) | (0.004) | (0.001) |
| Size of area | -0.000 | 0.000*** | -0.000*** | 0.000 | -0.000 | 0.000*** | -0.000*** | 0.000 |
| | (0.000) | (0.000) | (0.000) | (0.000) | (0.000) | (0.000) | (0.000) | (0.000) |
| Electricity | 0.012 | 0.028*** | 0.005 | -0.004*** | 0.005 | 0.028*** | 0.003 | -0.004*** |
| | (0.008) | (0.004) | (0.004) | (0.001) | (0.008) | (0.004) | (0.004) | (0.001) |
| Primary roads | 0.021*** | -0.000 | 0.004* | -0.000 | 0.026*** | -0.000 | 0.004* | 0.000 |
| | (0.004) | (0.002) | (0.002) | (0.001) | (0.004) | (0.002) | (0.002) | (0.001) |
| Log population | -0.026*** | -0.001 | 0.000 | 0.000 | -0.027*** | -0.001 | 0.000 | -0.000 |
| | (0.003) | (0.001) | (0.001) | (0.000) | (0.003) | (0.001) | (0.001) | (0.000) |
| Log infant mortality rate | 0.079*** | 0.001 | -0.018** | 0.006*** | 0.063*** | 0.001 | -0.027*** | 0.008*** |
| | (0.014) | (0.007) | (0.007) | (0.002) | (0.015) | (0.007) | (0.007) | (0.002) |
| Log cultivated | -0.027*** | -0.004* | -0.017*** | 0.001 | -0.031*** | -0.004* | -0.017*** | 0.002** |
| | (0.005) | (0.003) | (0.003) | (0.001) | (0.006) | (0.003) | (0.003) | (0.001) |
| Ethnolinguistic fractionalisation index | -0.015 | 0.058 | -0.019 | -0.007* | 0.037 | 0.065 | 0.000 | -0.005 |
| | (0.056) | (0.058) | (0.031) | (0.004) | (0.114) | (0.088) | (0.075) | (0.011) |
| Constant | 43.927*** | -17.621 | 10.271 | 1.157 | 20.167 | -21.776 | -2.595 | -0.004 |
| | (11.585) | (12.437) | (6.442) | (0.816) | (42.825) | (33.827) | (28.750) | (4.000) |
| Observations | 42,010 | 42,010 | 42,010 | 42,010 | 42,010 | 42,010 | 42,010 | 42,010 |
| Number of countries | 24 | 24 | 24 | 24 | 24 | 24 | 24 | 24 |
| Test of overidentification restrictions: | | | | | | | | |
| Sargan-Hanssen statistics Chi-sq(1) | | | | | 9.349 | 1.865 | 3.949 | 5.014 |
| P-value | | | | | 0.155 | 0.932 | 0.684 | 0.542 |
| Hausman test | | | | | | | | |
| Chi2 | | | | | 62.150 | 3.360 | 68.150 | 32.720 |
| Prob>chi2 | | | | | 0.000 | 1.000 | 0.000 | 0.012 |

The Sargan-Hanssen overidentification tests show that the instruments satisfy the overidentification restrictions. Also, the Hausman tests suggest the IV-2SLS specifications should be preferred. According to these specifications, since the local lockdowns, the instances where the state is involved in food-related conflicts has increased (column 8). However, we find that in countries that have provided a higher number of welfare and labour anti-poverty policies, the state is less likely to be involved as an actor in food-related conflicts. In contrast, in these countries the state is more likely to be involved as an actor in violence against civilians (column 7), but perhaps in ensuring lockdowns and preventing unrests.

## 6. CONCLUSION

We analysed the impact of social distancing measures, food vulnerability, welfare and labour COVID-19 policy response on conflict. Our IV-2SLS specifications revealed that despite the restrictions on population mobility, riots, violence against civilians and food-related increased after lockdowns. Food insecurity, in terms of volatility of local prices, was found to be associated with a higher probability of a country experiencing violence against civilians. Nonetheless, we also found that countries with a higher index of welfare and labour COVID-19 policy response are less likely to have suffered these conflicts and less likely to have experienced fatalities as a result of violence against civilians and any other conflicts. We also found that since the lockdown states have been more heavily involved as actors in food-related conflicts. However, states with higher welfare and labour COVID-19 policy index also are less likely to have to intervene in food-related conflicts directly.

      The implications of our analysis are important from a public policy perspective. Food vulnerability and price volatility are an explosive combination for conflicts as they provide an opportunity for rebel groups to attack civilians, particularly in areas with a high level of cultivation. Indeed, we found evidence that food vulnerability has increased the probability of experiencing violence against civilians. This evidence is well in line with the theoretical literature that suggests vulnerable citizens are more likely to join riots and fall prey to organised armed groups (Rezaeedaryakenari et al., 2020). However, our results also indicate that state's actions in terms of delivering urgent welfare assistance can reduce the probability of experiencing riots, violence against civilians, food-related conflicts as well as their associated casualties. Although the association found is weak, the findings are encouraging to suggest that urgent state interventions can reduce food vulnerability and prevent major social unrest.

Rezaeedaryakenari, B., Landis, S. T., & Thies, C. G. (2020). Food price volatilities and civilian victimization in Africa. *Conflict Management and Peace Science*, *37*(2), 193–214.

Senghore, M., Savi, M. K., Gnangnon, B., Hanage, W. P., & Okeke, I. N. (2020, May). Leveraging Africa's preparedness towards the next phase of the COVID-19 pandemic. *The Lancet Global Health*.

Sumner, A., Hoy, C., & Ortiz-Juarez, E. (2020). *Estimates of the impact of COVID-19 on global poverty.* (WIDER Working Paper No. 2020/43). Helsinki.

The Economist. (2020, May). Covid nostra - The pandemic is creating fresh opportunities for organised crime. *The Economist*.

Tiwari, S., Daidone, S., Ruvalcaba, M. A., Prifti, E., Handa, S., Davis, B., … Seidenfeld, D. (2016). Impact of cash transfer programs on food security and nutrition in sub-Saharan Africa: A cross-country analysis. *Global Food Security*, *11*, 72–83.

Tondo, L. (2020, April 10). Mafia distributes food to Italy's struggling residents. *The Guardian*.

UN News. (2020, March 23). COVID-19: UN chief calls for global ceasefire to focus on 'the true fight of our lives.' *2020*.

WFM. (2020, April 29). Futures of 370 million children in jeopardy as school closures deprive them of school meals – UNICEF and WFP.
33

APPENDIX

Table A.1. *Countries analysed with data on local food prices at sub-level until 2020*

| Country | Freq. | Percent | Date of first social distancing | Date of start of local lockdown |
|---|---|---|---|---|
| Algeria | 4,558 | 10.85 | 10-Mar-20 | 10-Mar-20 |
| Angola | 301 | 0.72 | 06-Feb-20 | 20-Mar-20 |
| Benin | 169 | 0.4 | 03-Mar-20 | 19-Mar-20 |
| Burkina Faso | 2,013 | 4.79 | 01-Jan-20 | 12-Mar-20 |
| Burundi | 5,525 | 13.15 | 06-Mar-20 | 12-Mar-20 |
| Cameroon | 2,619 | 6.23 | 01-Jan-20 | 18-Mar-20 |
| Central African Republic | 458 | 1.09 | 29-Jan-20 | 13-Mar-20 |
| Democratic Republic of Congo | 5,630 | 13.4 | 20-Feb-20 | 18-Mar-20 |
| Ethiopia | 1,389 | 3.31 | 16-Mar-20 | 16-Mar-20 |
| Gabon | 155 | 0.37 | 07-Feb-20 | 13-Mar-20 |
| Ghana | 715 | 1.7 | 24-Jan-20 | 16-Mar-20 |
| Guinea | 886 | 2.11 | 29-Feb-20 | 26-Mar-20 |
| Kenya | 2,528 | 6.02 | 20-Jan-20 | 13-Mar-20 |
| Lesotho | 39 | 0.09 | 06-Mar-20 | 18-Mar-20 |
| Liberia | 340 | 0.81 | 09-Mar-20 | 11-Apr-20 |
| Madagascar | 771 | 1.84 | 15-Mar-20 | 20-Mar-20 |
| Malawi | 405 | 0.96 | 16-Mar-20 | 16-Mar-20 |
| Mali | 1,206 | 2.87 | 19-Mar-20 | 19-Mar-20 |
| Mauritania | 42 | 0.1 | 05-Feb-20 | 16-Mar-20 |
| Namibia | 242 | 0.58 | 01-Mar-20 | 17-Mar-20 |
| Niger | 737 | 1.75 | 13-Mar-20 | 13-Mar-20 |
| Nigeria | 9,824 | 23.38 | 01-Jan-20 | 29/03/2020 |
| Rwanda | 93 | 0.22 | 27-Jan-20 | 08-Mar-20 |
| Zimbabwe | 1,365 | 3.25 | 28-Jan-20 | 17-Mar-20 |
| Total ACLED events | 42,010 | 100 | | |

Sources: Conflict events, ACLED. Dates on social distancing and lockdowns own estimates using ACAPS (2020) and Hale et al. (2020).

Table A.2. *Welfare and labour COVID-19 policy response of 24 countries analysed*

| | Overall COVID-19 index | SOCIAL ASSISTANCE | | | | SOCIAL INSURANCE | | | | LABOUR MARKETS | | | |
|---|---|---|---|---|---|---|---|---|---|---|---|---|---|
| | | Cash-Public based transfers | Public Works | In-kind (in-kind/school feeding) | Utility and financial support | Paid leave/ unemployment | Health insurance support | Pensions and disability benefits | Social security contributions (waiver/subsidy) | Wage waiver/subsidy | Activation (training) | Labour regulation adjustment | Reduced work time subsidy |
| Algeria | 0.417 | 1 | 0 | 1 | 0 | 1 | 0 | 1 | 1 | 0 | 0 | 0 | 0 |
| Angola | 0.083 | 1 | 0 | 0 | 0 | 0 | 0 | 0 | 0 | 0 | 0 | 0 | 0 |
| Benin | 0.083 | 0 | 0 | 0 | 1 | 0 | 0 | 0 | 0 | 0 | 0 | 0 | 0 |
| Burkina Faso | 0.250 | 1 | 0 | 1 | 1 | 0 | 0 | 0 | 0 | 0 | 0 | 0 | 0 |
| Burundi | 0.000 | - | - | - | - | - | - | - | - | - | - | - | - |
| Cameroon | 0.083 | 0 | 0 | 0 | 1 | 0 | 0 | 0 | 0 | 0 | 0 | 0 | 0 |
| Central African Republic | 0.000 | - | - | - | - | - | - | - | - | - | - | - | - |
| Democratic Republic of Congo | 0.000 | - | - | - | - | - | - | - | - | - | - | - | - |
| Ethiopia | 0.333 | 0 | 1 | 1 | 1 | 0 | 0 | 0 | 0 | 0 | 0 | 1 | 0 |
| Gabon | 0.000 | - | - | - | - | - | - | - | - | - | - | - | - |
| Ghana | 0.250 | 0 | 0 | 1 | 1 | 0 | 0 | 1 | 0 | 0 | 0 | 0 | 0 |
| Guinea | 0.167 | 1 | 0 | 1 | 0 | 0 | 0 | 0 | 0 | 0 | 0 | 0 | 0 |
| Kenya | 0.167 | 1 | 0 | 0 | 1 | 0 | 0 | 0 | 0 | 0 | 0 | 0 | 0 |
| Lesotho | 0.000 | - | - | - | - | - | - | - | - | - | - | - | - |
| Liberia | 0.167 | 0 | 0 | 1 | 1 | 0 | 0 | 0 | 0 | 0 | 0 | 0 | 0 |
| Madagascar | 0.250 | 1 | 0 | 1 | 0 | 0 | 0 | 0 | 1 | 0 | 0 | 0 | 0 |
| Malawi | 0.083 | 1 | 0 | 0 | 0 | 0 | 0 | 0 | 0 | 0 | 0 | 0 | 0 |
| Mali | 0.167 | 0 | 0 | 1 | 1 | 0 | 0 | 0 | 0 | 0 | 0 | 0 | 0 |
| Mauritania | 0.167 | 1 | 0 | 0 | 1 | 0 | 0 | 0 | 0 | 0 | 0 | 0 | 0 |
| Namibia | 0.167 | 1 | 0 | 0 | 1 | 0 | 0 | 0 | 0 | 0 | 0 | 0 | 0 |
| Niger | 0.083 | 0 | 0 | 0 | 1 | 0 | 0 | 0 | 0 | 0 | 0 | 0 | 0 |
| Nigeria | 0.250 | 1 | 0 | 1 | 1 | 0 | 0 | 0 | 0 | 0 | 0 | 0 | 0 |
| Rwanda | 0.333 | 1 | 0 | 1 | 1 | 0 | 0 | 0 | 1 | 0 | 0 | 0 | 0 |
| Zimbabwe | 0.083 | 1 | 0 | 0 | 0 | 0 | 0 | 0 | 0 | 0 | 0 | 0 | 0 |

Note: - No programme implemented until 1 May 2020. Source: Gentilini et al. (2020).

Table A.3. *Data sources*

| Variable | Source |
| --- | --- |
| All conflicts analysed, fatalities, and state involved as actor | Own construction using ACLED. |
| Date of social distancing and lockdowns | Own construction using Hale et al. (2020) and ACAPS (2020). |
| Index of welfare and labour COVID-19 response | Own construction using Gentilini et al. (2020). |
| Date of start of welfare/labour COVID-19 response | Own construction using Hale et al. (2020). |
| Index local market price | Own construction using the Global Food Prices Database (WFP) and for Zimbabwe only the USAID FEWS-NET. |
| Log stable nightlight (year 2015) | USA Air Force Weather Agency. |
| Cultivated land by district | Rezaeedaryakenari, Landis and Thies' (2020). Publicly available data. They used the Global Agro-Ecological Zones (GAEZ) of Food and Agricultu |
| Size of area (district) | Rezaeedaryakenari, Landis and Thies' (2020). Publicly available data. |
| Log mobile phone coverage 2G-3G | Manacorda and Tesei's (2020) publicly available data. They used the Global System for Mobile Communications (GSM) Association. |
| % Mountains | Manacorda and Tesei's (2020) publicly available data. They used UNEP-WCMC. |
| % Forests | Manacorda and Tesei's (2020) publicly available data. They used GLOBCover. |
| Petroleum fields | Manacorda and Tesei's (2020) publicly available data. They used PRIO. |
| Mines | Manacorda and Tesei's (2020) publicly available data. They used USA Geological Survey. |
| Diamond mines | Manacorda and Tesei's (2020) publicly available data. They used PRIO. |
| Electricity | Manacorda and Tesei's (2020) publicly available data. They used the Africa Infrastructure Country diagnostic (ADB). |
| Primary roads | Manacorda and Tesei's (2020) publicly available data. They used the Africa Infrastructure Country diagnostic (ADB). |
| Population | Manacorda and Tesei's (2020) publicly available data. They used SEDAC/NASA. |
| Log infant mortality rate | Manacorda and Tesei's (2020) publicly available data. They used SEDAC/NASA. |
| Ethnolinguistic fractionalisation index | Altas Maradov Mira |
| Male mortality rate attributed to household and ambient air pollution, age-standarised at national level, year 2016 | World Bank data repository |
| Adult diabetes prevalence (% of population ages 20 to 79) at national level | World Bank data repository |
| IMF global commodity price | IMF data repository |

Table A.4. *Summary statistics of countries analysed*

|  | 1 January 2015-6 May 2020 | | | 1 October-31 December 2019 | | | After lockdown in 2020 | | |
|---|---|---|---|---|---|---|---|---|---|
| Variable | Total | Mean | Std. Dev. | Total | Mean | Std. Dev. | Total | Mean | Std. Dev. |
| Riots | 12572 | 0.13 | 0.33 | 524 | 0.08 | 0.28 | 346 | 0.135 | 0.342 |
| Violence against civilians | 24745 | 0.28 | 0.45 | 1304 | 0.23 | 0.42 | 854 | 0.384 | 0.487 |
| Food-related incidents | 2871 | 0.02 | 0.16 | 174 | 0.03 | 0.17 | 160 | 0.047 | 0.211 |
| Food looting | 1798 | 0.02 | 0.12 | 110 | 0.02 | 0.13 | 107 | 0.026 | 0.160 |
| Fatalaties any ACLED conflict | 169454 | 1.66 | 8.59 | 6489 | 1.08 | 3.71 | 4616 | 1.894 | 6.172 |
| Fatalalties to riots | 4552 | 0.06 | 0.89 | 272 | 0.04 | 0.36 | 134 | 0.065 | 0.415 |
| Fatalaties to violence against civilians | 50506 | 0.69 | 6.37 | 1816 | 0.38 | 1.81 | 1236 | 0.583 | 2.360 |
| Fatalaties to food-related conflict | 6888 | 0.05 | 1.08 | 235 | 0.04 | 0.78 | 290 | 0.092 | 2.482 |
| Fatalaties to food looting | 4344 | 0.03 | 0.80 | 154 | 0.03 | 0.71 | 225 | 0.077 | 2.447 |
| State involved as actor in any ACLED conflict | 40237 | 0.32 | 0.47 | 2083 | 0.26 | 0.44 | 1548 | 0.404 | 0.491 |
| State involved as actor in riots | 4710 | 0.05 | 0.21 | 180 | 0.03 | 0.17 | 157 | 0.056 | 0.231 |
| State involved as actor in violence against civilians | 5309 | 0.05 | 0.22 | 225 | 0.03 | 0.17 | 279 | 0.114 | 0.318 |
| State involved as actor in food-related conflict | 691 | 0.01 | 0.07 | 23 | 0.00 | 0.06 | 41 | 0.011 | 0.106 |
| State involved as actor in food looting | 396 | 0.00 | 0.05 | 10 | 0.00 | 0.04 | 26 | 0.007 | 0.082 |
| Controls and instruments | | | | | | | | | |
| Log index local market price | | 4.82 | 0.49 | | 4.74 | 0.40 | | 4.768 | 0.411 |
| Adult diabetes prevalence (% of population ages 20 to 79) | | 4.26 | 1.73 | | 5.01 | 1.70 | | 4.696 | 1.739 |
| IMF global commodity price | | 113.80 | 11.68 | | 116.60 | 2.78 | | 86.988 | 4.413 |
| Log stable nightlight, year 2015 | | 1.92 | 0.72 | | | | | | |
| Log mobile phone coverage 2G-3G | | -0.52 | 0.93 | | | | | | |
| % Mountains | | 0.33 | 0.34 | | | | | | |
| % Forests | | 0.24 | 0.22 | | | | | | |
| Petroleum fields | | 0.06 | 0.20 | | | | | | |
| Mines | | 0.30 | 0.63 | | | | | | |
| Diamond mines | | 0.04 | 0.32 | | | | | | |
| Size of area | | 2989.24 | 613.91 | | | | | | |
| Electricity | | 0.44 | 0.44 | | | | | | |
| Primary roads | | 1.88 | 1.66 | | | | | | |
| Log population | | 12.86 | 1.39 | | | | | | |
| Log infant mortality rate | | 2.11 | 0.43 | | | | | | |
| Log cultivated | | 3.89 | 0.65 | | | | | | |
| Ethnolinguistic fractionalisation index | | 0.61 | 0.29 | | | | | | |
| Index of welfare and labour COVID-19 response | | 0.01 | 0.04 | | | | | | |
| Male mortality rate attributed to household and ambient air pollution, age-standarised, year 2016 | | 192.60 | 79.43 | | | | | | |
| Number of observations | 42010 | | | 3134 | | | 1330 | | |
| Number of countries | 24 | | | 24 | | | 24 | | |

Table A.5. *First-stage regression of Table 1, COVID-interventions and conflict*

|  | Riots | | | Violence against civilians | | | Food-related incidents | | | Food looting | | |
|---|---|---|---|---|---|---|---|---|---|---|---|---|
|  | (1) | (2) | (3) | (4) | (5) | (6) | (7) | (8) | (9) | (10) | (11) | (12) |
|  | First social distancing | Strict lockdown | Index welfare/ labour | First social distancing | Strict lockdown | Index welfare/ labour | First social distancing | Strict lockdown | Index welfare/ labour | First social distancing | Strict lockdown | Index welfare/ labour |
| Male mortality rate attributed to household and ambient air pollution male | -0.120*** | 0.000** | -0.000*** | -0.120*** | 0.000** | -0.000*** | -0.120*** | 0.000** | -0.000*** | -0.120*** | 0.000** | -0.000*** |
|  | (0.001) | (0.000) | (0.000) | (0.001) | (0.000) | (0.000) | (0.001) | (0.000) | (0.000) | (0.001) | (0.000) | (0.000) |
| Diabetes prevalence (% of population ages 20 to 79) | -3.189*** | 0.005*** | -0.003*** | -3.189*** | 0.005*** | -0.003*** | -3.189*** | 0.005*** | -0.003*** | -3.189*** | 0.005*** | -0.003*** |
|  | (0.054) | (0.001) | (0.000) | (0.054) | (0.001) | (0.000) | (0.054) | (0.001) | (0.000) | (0.054) | (0.001) | (0.000) |
| Former colony (never colonised reference group): | | | | | | | | | | | | |
| British | -43.649*** | 0.037*** | 0.009*** | -43.649*** | 0.037*** | 0.009*** | -43.649*** | 0.037*** | 0.009*** | -43.649*** | 0.037*** | 0.009*** |
|  | (0.480) | (0.006) | (0.001) | (0.480) | (0.006) | (0.001) | (0.480) | (0.006) | (0.001) | (0.480) | (0.006) | (0.001) |
| French | -14.998*** | 0.069*** | 0.020*** | -14.998*** | 0.069*** | 0.020*** | -14.998*** | 0.069*** | 0.020*** | -14.998*** | 0.069*** | 0.020*** |
|  | (0.476) | (0.006) | (0.001) | (0.476) | (0.006) | (0.001) | (0.476) | (0.006) | (0.001) | (0.476) | (0.006) | (0.001) |
| Portuguese | -36.827*** | 0.037*** | 0.013*** | -36.827*** | 0.037*** | 0.013*** | -36.827*** | 0.037*** | 0.013*** | -36.827*** | 0.037*** | 0.013*** |
|  | (0.892) | (0.011) | (0.002) | (0.892) | (0.011) | (0.002) | (0.892) | (0.011) | (0.002) | (0.892) | (0.011) | (0.002) |
| German | -45.109*** | 0.063*** | 0.007*** | -45.109*** | 0.063*** | 0.007*** | -45.109*** | 0.063*** | 0.007*** | -45.109*** | 0.063*** | 0.007*** |
|  | (0.554) | (0.007) | (0.002) | (0.554) | (0.007) | (0.002) | (0.554) | (0.007) | (0.002) | (0.554) | (0.007) | (0.002) |
| Belgium | -16.255*** | 0.049*** | 0.012*** | -16.255*** | 0.049*** | 0.012*** | -16.255*** | 0.049*** | 0.012*** | -16.255*** | 0.049*** | 0.012*** |
|  | (0.485) | (0.006) | (0.001) | (0.485) | (0.006) | (0.001) | (0.485) | (0.006) | (0.001) | (0.485) | (0.006) | (0.001) |
| American Colonisation Society | 12.062*** | 0.023** | 0.020*** | 12.062*** | 0.023** | 0.020*** | 12.062*** | 0.023** | 0.020*** | 12.062*** | 0.023** | 0.020*** |
|  | (0.876) | (0.011) | (0.002) | (0.876) | (0.011) | (0.002) | (0.876) | (0.011) | (0.002) | (0.876) | (0.011) | (0.002) |
| IMF all commodity price | -0.021*** | -0.006*** | -0.001*** | -0.021*** | -0.006*** | -0.001*** | -0.021*** | -0.006*** | -0.001*** | -0.021*** | -0.006*** | -0.001*** |
|  | (0.006) | (0.000) | (0.000) | (0.006) | (0.000) | (0.000) | (0.006) | (0.000) | (0.000) | (0.006) | (0.000) | (0.000) |
| Log index local market price | 1.497*** | 0.017*** | 0.001*** | 1.497*** | 0.017*** | 0.001*** | 1.497*** | 0.017*** | 0.001*** | 1.497*** | 0.017*** | 0.001*** |
|  | (0.160) | (0.002) | (0.000) | (0.160) | (0.002) | (0.000) | (0.160) | (0.002) | (0.000) | (0.160) | (0.002) | (0.000) |
| Log stable nightlight (year 2015) | 4.162*** | -0.011*** | -0.000 | 4.162*** | -0.011*** | -0.000 | 4.162*** | -0.011*** | -0.000 | 4.162*** | -0.011*** | -0.000 |
|  | (0.139) | (0.002) | (0.000) | (0.139) | (0.002) | (0.000) | (0.139) | (0.002) | (0.000) | (0.139) | (0.002) | (0.000) |
| Log mobile phone coverage 2G-3G | -2.230*** | 0.002 | -0.000 | -2.230*** | 0.002 | -0.000 | -2.230*** | 0.002 | -0.000 | -2.230*** | 0.002 | -0.000 |
|  | (0.084) | (0.001) | (0.000) | (0.084) | (0.001) | (0.000) | (0.084) | (0.001) | (0.000) | (0.084) | (0.001) | (0.000) |
| % Mountains | 6.402*** | 0.006* | 0.000 | 6.402*** | 0.006* | 0.000 | 6.402*** | 0.006* | 0.000 | 6.402*** | 0.006* | 0.000 |
|  | (0.294) | (0.004) | (0.001) | (0.294) | (0.004) | (0.001) | (0.294) | (0.004) | (0.001) | (0.294) | (0.004) | (0.001) |
| % Forests | -7.042*** | -0.003 | -0.004*** | -7.042*** | -0.003 | -0.004*** | -7.042*** | -0.003 | -0.004*** | -7.042*** | -0.003 | -0.004*** |
|  | (0.356) | (0.004) | (0.001) | (0.356) | (0.004) | (0.001) | (0.356) | (0.004) | (0.001) | (0.356) | (0.004) | (0.001) |
| Petroleum fields | 6.402*** | -0.011** | 0.001 | 6.402*** | -0.011** | 0.001 | 6.402*** | -0.011** | 0.001 | 6.402*** | -0.011** | 0.001 |
|  | (0.361) | (0.004) | (0.001) | (0.361) | (0.004) | (0.001) | (0.361) | (0.004) | (0.001) | (0.361) | (0.004) | (0.001) |
| Mines | 1.129*** | 0.006*** | 0.001** | 1.129*** | 0.006*** | 0.001** | 1.129*** | 0.006*** | 0.001** | 1.129*** | 0.006*** | 0.001** |
|  | (0.118) | (0.001) | (0.000) | (0.118) | (0.001) | (0.000) | (0.118) | (0.001) | (0.000) | (0.118) | (0.001) | (0.000) |
| Diamond mines | 2.604*** | -0.001 | 0.001** | 2.604*** | -0.001 | 0.001** | 2.604*** | -0.001 | 0.001** | 2.604*** | -0.001 | 0.001** |
|  | (0.207) | (0.002) | (0.001) | (0.207) | (0.002) | (0.001) | (0.207) | (0.002) | (0.001) | (0.207) | (0.002) | (0.001) |
| Size of area | -0.000 | -0.000 | -0.000 | -0.000 | -0.000 | -0.000 | -0.000 | -0.000 | -0.000 | -0.000 | -0.000 | -0.000 |
|  | (0.000) | (0.000) | (0.000) | (0.000) | (0.000) | (0.000) | (0.000) | (0.000) | (0.000) | (0.000) | (0.000) | (0.000) |
| Electricity | -3.413*** | -0.002 | -0.002*** | -3.413*** | -0.002 | -0.002*** | -3.413*** | -0.002 | -0.002*** | -3.413*** | -0.002 | -0.002*** |
|  | (0.202) | (0.002) | (0.001) | (0.202) | (0.002) | (0.001) | (0.202) | (0.002) | (0.001) | (0.202) | (0.002) | (0.001) |
| Primary roads | 1.498*** | -0.003*** | -0.001*** | 1.498*** | -0.003*** | -0.001*** | 1.498*** | -0.003*** | -0.001*** | 1.498*** | -0.003*** | -0.001*** |
|  | (0.075) | (0.001) | (0.000) | (0.075) | (0.001) | (0.000) | (0.075) | (0.001) | (0.000) | (0.075) | (0.001) | (0.000) |
| Log population | -2.710*** | 0.002*** | 0.001*** | -2.710*** | 0.002*** | 0.001*** | -2.710*** | 0.002*** | 0.001*** | -2.710*** | 0.002*** | 0.001*** |
|  | (0.074) | (0.001) | (0.000) | (0.074) | (0.001) | (0.000) | (0.074) | (0.001) | (0.000) | (0.074) | (0.001) | (0.000) |
| Log infant mortality rate | -0.279 | 0.012*** | -0.015*** | -0.279 | 0.012*** | -0.015*** | -0.279 | 0.012*** | -0.015*** | -0.279 | 0.012*** | -0.015*** |
|  | (0.282) | (0.003) | (0.001) | (0.282) | (0.003) | (0.001) | (0.282) | (0.003) | (0.001) | (0.282) | (0.003) | (0.001) |
| Log cultivated | 3.223*** | 0.004** | 0.003*** | 3.223*** | 0.004** | 0.003*** | 3.223*** | 0.004** | 0.003*** | 3.223*** | 0.004** | 0.003*** |
|  | (0.137) | (0.002) | (0.000) | (0.137) | (0.002) | (0.000) | (0.137) | (0.002) | (0.000) | (0.137) | (0.002) | (0.000) |
| Ethnolinguistic fractionalisation index | -19.392*** | 0.036*** | 0.011*** | -19.392*** | 0.036*** | 0.011*** | -19.392*** | 0.036*** | 0.011*** | -19.392*** | 0.036*** | 0.011*** |
|  | (0.499) | (0.006) | (0.001) | (0.499) | (0.006) | (0.001) | (0.499) | (0.006) | (0.001) | (0.499) | (0.006) | (0.001) |
| Observations | 42,010 | 42,010 | 42,010 | 42,010 | 42,010 | 42,010 | 42,010 | 42,010 | 42,010 | 42,010 | 42,010 | 42,010 |
| R-squared | 0.817 | 0.186 | 0.113 | 0.817 | 0.186 | 0.113 | 0.817 | 0.186 | 0.113 | 0.817 | 0.186 | 0.113 |

Table A.6. *First-stage regression of Table 2, COVID-interventions and fatalities*

| Fatalaties of: | Any ACLED conflict | | | Riots | | | Violence against civilians | | | Food-related conflict | | |
|---|---|---|---|---|---|---|---|---|---|---|---|---|
| | (1) | (2) | (3) | (4) | (5) | (6) | (7) | (8) | (9) | (10) | (11) | (12) |
| | First social distancing | Strict lockdown | Index welfare/ labour | First social distancing | Strict lockdown | Index welfare/ labour | First social distancing | Strict lockdown | Index welfare/ labour | First social distancing | Strict lockdown | Index welfare/ labour |
| Male mortality rate attributed to household and ambient air pollution male | -0.120*** | 0.000** | -0.000*** | -0.120*** | 0.000** | -0.000*** | -0.120*** | 0.000** | -0.000*** | -0.120*** | 0.000** | -0.000*** |
| | (0.001) | (0.000) | (0.000) | (0.001) | (0.000) | (0.000) | (0.001) | (0.000) | (0.000) | (0.001) | (0.000) | (0.000) |
| Diabetes prevalence (% of population ages 20 to 79) | -3.189*** | 0.005*** | -0.003*** | -3.189*** | 0.005*** | -0.003*** | -3.189*** | 0.005*** | -0.003*** | -3.189*** | 0.005*** | -0.003*** |
| | (0.054) | (0.001) | (0.000) | (0.054) | (0.001) | (0.000) | (0.054) | (0.001) | (0.000) | (0.054) | (0.001) | (0.000) |
| Former colony (never colonised reference group): | | | | | | | | | | | | |
| British | -43.649*** | 0.037*** | 0.009*** | -43.649*** | 0.037*** | 0.009*** | -43.649*** | 0.037*** | 0.009*** | -43.649*** | 0.037*** | 0.009*** |
| | (0.480) | (0.006) | (0.001) | (0.480) | (0.006) | (0.001) | (0.480) | (0.006) | (0.001) | (0.480) | (0.006) | (0.001) |
| French | -14.998*** | 0.069*** | 0.020*** | -14.998*** | 0.069*** | 0.020*** | -14.998*** | 0.069*** | 0.020*** | -14.998*** | 0.069*** | 0.020*** |
| | (0.476) | (0.006) | (0.001) | (0.476) | (0.006) | (0.001) | (0.476) | (0.006) | (0.001) | (0.476) | (0.006) | (0.001) |
| Portuguese | -36.827*** | 0.037*** | 0.013*** | -36.827*** | 0.037*** | 0.013*** | -36.827*** | 0.037*** | 0.013*** | -36.827*** | 0.037*** | 0.013*** |
| | (0.892) | (0.011) | (0.002) | (0.892) | (0.011) | (0.002) | (0.892) | (0.011) | (0.002) | (0.892) | (0.011) | (0.002) |
| German | -45.109*** | 0.063*** | 0.007*** | -45.109*** | 0.063*** | 0.007*** | -45.109*** | 0.063*** | 0.007*** | -45.109*** | 0.063*** | 0.007*** |
| | (0.554) | (0.007) | (0.002) | (0.554) | (0.007) | (0.002) | (0.554) | (0.007) | (0.002) | (0.554) | (0.007) | (0.002) |
| Belgium | -16.255*** | 0.049*** | 0.012*** | -16.255*** | 0.049*** | 0.012*** | -16.255*** | 0.049*** | 0.012*** | -16.255*** | 0.049*** | 0.012*** |
| | (0.485) | (0.006) | (0.001) | (0.485) | (0.006) | (0.001) | (0.485) | (0.006) | (0.001) | (0.485) | (0.006) | (0.001) |
| American Colonisation Society | 12.062*** | 0.023** | 0.020*** | 12.062*** | 0.023** | 0.020*** | 12.062*** | 0.023** | 0.020*** | 12.062*** | 0.023** | 0.020*** |
| | (0.876) | (0.011) | (0.002) | (0.876) | (0.011) | (0.002) | (0.876) | (0.011) | (0.002) | (0.876) | (0.011) | (0.002) |
| IMF all commodity price | -0.021*** | -0.006*** | -0.001*** | -0.021*** | -0.006*** | -0.001*** | -0.021*** | -0.006*** | -0.001*** | -0.021*** | -0.006*** | -0.001*** |
| | (0.006) | (0.000) | (0.000) | (0.006) | (0.000) | (0.000) | (0.006) | (0.000) | (0.000) | (0.006) | (0.000) | (0.000) |
| Log index local market price | 1.497*** | 0.017*** | 0.001*** | 1.497*** | 0.017*** | 0.001*** | 1.497*** | 0.017*** | 0.001*** | 1.497*** | 0.017*** | 0.001*** |
| | (0.160) | (0.002) | (0.000) | (0.160) | (0.002) | (0.000) | (0.160) | (0.002) | (0.000) | (0.160) | (0.002) | (0.000) |
| Log stable nightlight (year 2015) | 4.162*** | -0.011*** | -0.000 | 4.162*** | -0.011*** | -0.000 | 4.162*** | -0.011*** | -0.000 | 4.162*** | -0.011*** | -0.000 |
| | (0.139) | (0.002) | (0.000) | (0.139) | (0.002) | (0.000) | (0.139) | (0.002) | (0.000) | (0.139) | (0.002) | (0.000) |
| Log mobile phone coverage 2G-3G | -2.230*** | 0.002 | -0.000 | -2.230*** | 0.002 | -0.000 | -2.230*** | 0.002 | -0.000 | -2.230*** | 0.002 | -0.000 |
| | (0.084) | (0.001) | (0.000) | (0.084) | (0.001) | (0.000) | (0.084) | (0.001) | (0.000) | (0.084) | (0.001) | (0.000) |
| % Mountains | 6.402*** | 0.006* | 0.000 | 6.402*** | 0.006* | 0.000 | 6.402*** | 0.006* | 0.000 | 6.402*** | 0.006* | 0.000 |
| | (0.294) | (0.004) | (0.001) | (0.294) | (0.004) | (0.001) | (0.294) | (0.004) | (0.001) | (0.294) | (0.004) | (0.001) |
| % Forests | -7.042*** | -0.003 | -0.004*** | -7.042*** | -0.003 | -0.004*** | -7.042*** | -0.003 | -0.004*** | -7.042*** | -0.003 | -0.004*** |
| | (0.356) | (0.004) | (0.001) | (0.356) | (0.004) | (0.001) | (0.356) | (0.004) | (0.001) | (0.356) | (0.004) | (0.001) |
| Petroleum fields | 6.402*** | -0.011** | 0.001 | 6.402*** | -0.011** | 0.001 | 6.402*** | -0.011** | 0.001 | 6.402*** | -0.011** | 0.001 |
| | (0.361) | (0.004) | (0.001) | (0.361) | (0.004) | (0.001) | (0.361) | (0.004) | (0.001) | (0.361) | (0.004) | (0.001) |
| Mines | 1.129*** | 0.006*** | 0.001** | 1.129*** | 0.006*** | 0.001** | 1.129*** | 0.006*** | 0.001** | 1.129*** | 0.006*** | 0.001** |
| | (0.118) | (0.001) | (0.000) | (0.118) | (0.001) | (0.000) | (0.118) | (0.001) | (0.000) | (0.118) | (0.001) | (0.000) |
| Diamond mines | 2.604*** | -0.001 | 0.001** | 2.604*** | -0.001 | 0.001** | 2.604*** | -0.001 | 0.001** | 2.604*** | -0.001 | 0.001** |
| | (0.207) | (0.002) | (0.001) | (0.207) | (0.002) | (0.001) | (0.207) | (0.002) | (0.001) | (0.207) | (0.002) | (0.001) |
| Size of area | -0.000 | -0.000 | -0.000 | -0.000 | -0.000 | -0.000 | -0.000 | -0.000 | -0.000 | -0.000 | -0.000 | -0.000 |
| | (0.000) | (0.000) | (0.000) | (0.000) | (0.000) | (0.000) | (0.000) | (0.000) | (0.000) | (0.000) | (0.000) | (0.000) |
| Electricity | -3.413*** | -0.002 | -0.002*** | -3.413*** | -0.002 | -0.002*** | -3.413*** | -0.002 | -0.002*** | -3.413*** | -0.002 | -0.002*** |
| | (0.202) | (0.002) | (0.001) | (0.202) | (0.002) | (0.001) | (0.202) | (0.002) | (0.001) | (0.202) | (0.002) | (0.001) |
| Primary roads | 1.498*** | -0.003*** | -0.001*** | 1.498*** | -0.003*** | -0.001*** | 1.498*** | -0.003*** | -0.001*** | 1.498*** | -0.003*** | -0.001*** |
| | (0.075) | (0.001) | (0.000) | (0.075) | (0.001) | (0.000) | (0.075) | (0.001) | (0.000) | (0.075) | (0.001) | (0.000) |
| Log population | -2.710*** | 0.002*** | 0.001*** | -2.710*** | 0.002*** | 0.001*** | -2.710*** | 0.002*** | 0.001*** | -2.710*** | 0.002*** | 0.001*** |
| | (0.074) | (0.001) | (0.000) | (0.074) | (0.001) | (0.000) | (0.074) | (0.001) | (0.000) | (0.074) | (0.001) | (0.000) |
| Log infant mortality rate | -0.279 | 0.012*** | -0.015*** | -0.279 | 0.012*** | -0.015*** | -0.279 | 0.012*** | -0.015*** | -0.279 | 0.012*** | -0.015*** |
| | (0.282) | (0.003) | (0.001) | (0.282) | (0.003) | (0.001) | (0.282) | (0.003) | (0.001) | (0.282) | (0.003) | (0.001) |
| Log cultivated | 3.223*** | 0.004** | 0.003*** | 3.223*** | 0.004** | 0.003*** | 3.223*** | 0.004** | 0.003*** | 3.223*** | 0.004** | 0.003*** |
| | (0.137) | (0.002) | (0.000) | (0.137) | (0.002) | (0.000) | (0.137) | (0.002) | (0.000) | (0.137) | (0.002) | (0.000) |
| Ethnolinguistic fractionalisation index | -19.392*** | 0.036*** | 0.011*** | -19.392*** | 0.036*** | 0.011*** | -19.392*** | 0.036*** | 0.011*** | -19.392*** | 0.036*** | 0.011*** |
| | (0.499) | (0.006) | (0.001) | (0.499) | (0.006) | (0.001) | (0.499) | (0.006) | (0.001) | (0.499) | (0.006) | (0.001) |
| Observations | 42,010 | 42,010 | 42,010 | 42,010 | 42,010 | 42,010 | 42,010 | 42,010 | 42,010 | 42,010 | 42,010 | 42,010 |
| R-squared | 0.817 | 0.186 | 0.113 | 0.817 | 0.186 | 0.113 | 0.817 | 0.186 | 0.113 | 0.817 | 0.186 | 0.113 |

Table A.7. *First-stage regression of Table 3, COVID-interventions and the state*

| State (military, policy, gard or government) involved as actor in: | Any ACLED conflict | | | Riots | | | Violence against civilians | | | Food-related conflict | | |
|---|---|---|---|---|---|---|---|---|---|---|---|---|
| | (1) | (2) | (3) | (4) | (5) | (6) | (7) | (8) | (9) | (10) | (11) | (12) |
| | First social distancing | Strict lockdown | Index welfare/ labour | First social distancing | Strict lockdown | Index welfare/ labour | First social distancing | Strict lockdown | Index welfare/ labour | First social distancing | Strict lockdown | Index welfare/ labour |
| Male mortality rate attributed to household and ambient air pollution male | -0.120*** | 0.000** | -0.000*** | -0.120*** | 0.000** | -0.000*** | -0.120*** | 0.000** | -0.000*** | -0.120*** | 0.000** | -0.000*** |
| | (0.001) | (0.000) | (0.000) | (0.001) | (0.000) | (0.000) | (0.001) | (0.000) | (0.000) | (0.001) | (0.000) | (0.000) |
| Diabetes prevalence (% of population ages 20 to 79) | -3.189*** | 0.005*** | -0.003*** | -3.189*** | 0.005*** | -0.003*** | -3.189*** | 0.005*** | -0.003*** | -3.189*** | 0.005*** | -0.003*** |
| | (0.054) | (0.001) | (0.000) | (0.054) | (0.001) | (0.000) | (0.054) | (0.001) | (0.000) | (0.054) | (0.001) | (0.000) |
| Former colony (never colonised reference group): | | | | | | | | | | | | |
| British | -43.649*** | 0.037*** | 0.009*** | -43.649*** | 0.037*** | 0.009*** | -43.649*** | 0.037*** | 0.009*** | -43.649*** | 0.037*** | 0.009*** |
| | (0.480) | (0.006) | (0.001) | (0.480) | (0.006) | (0.001) | (0.480) | (0.006) | (0.001) | (0.480) | (0.006) | (0.001) |
| French | -14.998*** | 0.069*** | 0.020*** | -14.998*** | 0.069*** | 0.020*** | -14.998*** | 0.069*** | 0.020*** | -14.998*** | 0.069*** | 0.020*** |
| | (0.476) | (0.006) | (0.001) | (0.476) | (0.006) | (0.001) | (0.476) | (0.006) | (0.001) | (0.476) | (0.006) | (0.001) |
| Portuguese | -36.827*** | 0.037*** | 0.013*** | -36.827*** | 0.037*** | 0.013*** | -36.827*** | 0.037*** | 0.013*** | -36.827*** | 0.037*** | 0.013*** |
| | (0.892) | (0.011) | (0.002) | (0.892) | (0.011) | (0.002) | (0.892) | (0.011) | (0.002) | (0.892) | (0.011) | (0.002) |
| German | -45.109*** | 0.063*** | 0.007*** | -45.109*** | 0.063*** | 0.007*** | -45.109*** | 0.063*** | 0.007*** | -45.109*** | 0.063*** | 0.007*** |
| | (0.554) | (0.007) | (0.002) | (0.554) | (0.007) | (0.002) | (0.554) | (0.007) | (0.002) | (0.554) | (0.007) | (0.002) |
| Belgium | -16.255*** | 0.049*** | 0.012*** | -16.255*** | 0.049*** | 0.012*** | -16.255*** | 0.049*** | 0.012*** | -16.255*** | 0.049*** | 0.012*** |
| | (0.485) | (0.006) | (0.001) | (0.485) | (0.006) | (0.001) | (0.485) | (0.006) | (0.001) | (0.485) | (0.006) | (0.001) |
| American Colonisation Society | 12.062*** | 0.023** | 0.020*** | 12.062*** | 0.023** | 0.020*** | 12.062*** | 0.023** | 0.020*** | 12.062*** | 0.023** | 0.020*** |
| | (0.876) | (0.011) | (0.002) | (0.876) | (0.011) | (0.002) | (0.876) | (0.011) | (0.002) | (0.876) | (0.011) | (0.002) |
| IMF all commodity price | -0.021*** | -0.006*** | -0.001*** | -0.021*** | -0.006*** | -0.001*** | -0.021*** | -0.006*** | -0.001*** | -0.021*** | -0.006*** | -0.001*** |
| | (0.006) | (0.000) | (0.000) | (0.006) | (0.000) | (0.000) | (0.006) | (0.000) | (0.000) | (0.006) | (0.000) | (0.000) |
| Log index local market price | 1.497*** | 0.017*** | 0.001*** | 1.497*** | 0.017*** | 0.001*** | 1.497*** | 0.017*** | 0.001*** | 1.497*** | 0.017*** | 0.001*** |
| | (0.160) | (0.002) | (0.000) | (0.160) | (0.002) | (0.000) | (0.160) | (0.002) | (0.000) | (0.160) | (0.002) | (0.000) |
| Log stable nightlight (year 2015) | 4.162*** | -0.011*** | -0.000 | 4.162*** | -0.011*** | -0.000 | 4.162*** | -0.011*** | -0.000 | 4.162*** | -0.011*** | -0.000 |
| | (0.139) | (0.002) | (0.000) | (0.139) | (0.002) | (0.000) | (0.139) | (0.002) | (0.000) | (0.139) | (0.002) | (0.000) |
| Log mobile phone coverage 2G-3G | -2.230*** | 0.002 | -0.000 | -2.230*** | 0.002 | -0.000 | -2.230*** | 0.002 | -0.000 | -2.230*** | 0.002 | -0.000 |
| | (0.084) | (0.001) | (0.000) | (0.084) | (0.001) | (0.000) | (0.084) | (0.001) | (0.000) | (0.084) | (0.001) | (0.000) |
| % Mountains | 6.402*** | 0.006* | 0.000 | 6.402*** | 0.006* | 0.000 | 6.402*** | 0.006* | 0.000 | 6.402*** | 0.006* | 0.000 |
| | (0.294) | (0.004) | (0.001) | (0.294) | (0.004) | (0.001) | (0.294) | (0.004) | (0.001) | (0.294) | (0.004) | (0.001) |
| % Forests | -7.042*** | -0.003 | -0.004*** | -7.042*** | -0.003 | -0.004*** | -7.042*** | -0.003 | -0.004*** | -7.042*** | -0.003 | -0.004*** |
| | (0.356) | (0.004) | (0.001) | (0.356) | (0.004) | (0.001) | (0.356) | (0.004) | (0.001) | (0.356) | (0.004) | (0.001) |
| Petroleum fields | 6.402*** | -0.011** | 0.001 | 6.402*** | -0.011** | 0.001 | 6.402*** | -0.011** | 0.001 | 6.402*** | -0.011** | 0.001 |
| | (0.361) | (0.004) | (0.001) | (0.361) | (0.004) | (0.001) | (0.361) | (0.004) | (0.001) | (0.361) | (0.004) | (0.001) |
| Mines | 1.129*** | 0.006*** | 0.001** | 1.129*** | 0.006*** | 0.001** | 1.129*** | 0.006*** | 0.001** | 1.129*** | 0.006*** | 0.001** |
| | (0.118) | (0.001) | (0.000) | (0.118) | (0.001) | (0.000) | (0.118) | (0.001) | (0.000) | (0.118) | (0.001) | (0.000) |
| Diamond mines | 2.604*** | -0.001 | 0.001** | 2.604*** | -0.001 | 0.001** | 2.604*** | -0.001 | 0.001** | 2.604*** | -0.001 | 0.001** |
| | (0.207) | (0.002) | (0.001) | (0.207) | (0.002) | (0.001) | (0.207) | (0.002) | (0.001) | (0.207) | (0.002) | (0.001) |
| Size of area | -0.000 | -0.000 | -0.000 | -0.000 | -0.000 | -0.000 | -0.000 | -0.000 | -0.000 | -0.000 | -0.000 | -0.000 |
| | (0.000) | (0.000) | (0.000) | (0.000) | (0.000) | (0.000) | (0.000) | (0.000) | (0.000) | (0.000) | (0.000) | (0.000) |
| Electricity | -3.413*** | -0.002 | -0.002*** | -3.413*** | -0.002 | -0.002*** | -3.413*** | -0.002 | -0.002*** | -3.413*** | -0.002 | -0.002*** |
| | (0.202) | (0.002) | (0.001) | (0.202) | (0.002) | (0.001) | (0.202) | (0.002) | (0.001) | (0.202) | (0.002) | (0.001) |
| Primary roads | 1.498*** | -0.003*** | -0.001*** | 1.498*** | -0.003*** | -0.001*** | 1.498*** | -0.003*** | -0.001*** | 1.498*** | -0.003*** | -0.001*** |
| | (0.075) | (0.001) | (0.000) | (0.075) | (0.001) | (0.000) | (0.075) | (0.001) | (0.000) | (0.075) | (0.001) | (0.000) |
| Log population | -2.710*** | 0.002*** | 0.001*** | -2.710*** | 0.002*** | 0.001*** | -2.710*** | 0.002*** | 0.001*** | -2.710*** | 0.002*** | 0.001*** |
| | (0.074) | (0.001) | (0.000) | (0.074) | (0.001) | (0.000) | (0.074) | (0.001) | (0.000) | (0.074) | (0.001) | (0.000) |
| Log infant mortality rate | -0.279 | 0.012*** | -0.015*** | -0.279 | 0.012*** | -0.015*** | -0.279 | 0.012*** | -0.015*** | -0.279 | 0.012*** | -0.015*** |
| | (0.282) | (0.003) | (0.001) | (0.282) | (0.003) | (0.001) | (0.282) | (0.003) | (0.001) | (0.282) | (0.003) | (0.001) |
| Log cultivated | 3.223*** | 0.004** | 0.003*** | 3.223*** | 0.004** | 0.003*** | 3.223*** | 0.004** | 0.003*** | 3.223*** | 0.004** | 0.003*** |
| | (0.137) | (0.002) | (0.000) | (0.137) | (0.002) | (0.000) | (0.137) | (0.002) | (0.000) | (0.137) | (0.002) | (0.000) |
| Ethnolinguistic fractionalisation index | -19.392*** | 0.036*** | 0.011*** | -19.392*** | 0.036*** | 0.011*** | -19.392*** | 0.036*** | 0.011*** | -19.392*** | 0.036*** | 0.011*** |
| | (0.499) | (0.006) | (0.001) | (0.499) | (0.006) | (0.001) | (0.499) | (0.006) | (0.001) | (0.499) | (0.006) | (0.001) |
| Observations | 42,010 | 42,010 | 42,010 | 42,010 | 42,010 | 42,010 | 42,010 | 42,010 | 42,010 | 42,010 | 42,010 | 42,010 |
| R-squared | 0.817 | 0.186 | 0.113 | 0.817 | 0.186 | 0.113 | 0.817 | 0.186 | 0.113 | 0.817 | 0.186 | 0.113 |